\def\al{\alpha}
\def\be{\beta}
\def\ga{\gamma}
\def\de{\delta}
\def\ep{\epsilon}
\def\vep{\varepsilon}
\def\et{\eta}

\def\om{\omega}
\def\ps{\psi}

\def\De{\Delta}

\def\Om{\Omega}

\def\fr{\frac}

\def\pd{\partial}

\def\cd{\cdot}
\def\nb{\nabla}
\def\bv{\mbox{\boldmath $v$}}
\def\ra{\rightarrow}

\def\Bl{\Bigl}
\def\Br{\Bigr}

\newcommand{\Eq}[1]{(\ref{#1})}
\newcommand{\sq}[1]{\left[ #1\right]}
\newcommand{\Del}[1]{\Delta_{#1}}
\newcommand{\Ln}[1]{L_{#1}}
\newcommand{\Lnb}[1]{{\bar L}_{#1}}
\newcommand{\Pa}[2]{\frac{\partial #1}{\partial #2}}
\newcommand{\non}{\nonumber}
\newcommand{\zb}{\bar z}

\newcommand{\nc}{\newcommand}
\nc{\ellb}{\bar\ell}
\nc{\Lb}{\bar L}
\nc{\cmmt}[1]{\left[#1\right]}
\nc{\Deltab}{\bar\Delta}
\nc{\wb}{\bar w}

\documentstyle[11pt]{article}

\begin{document}

\begin{flushleft}
October 1993
\end{flushleft}
\begin{flushright}
UT-652
\end{flushright}

\large 
\centerline{ {\bf 
Predictions on Two-dimensional Turbulence 
by Conformal Field Theory } }

\ \ \

\normalsize
\centerline{ H. C$\hat{\mbox{a}}$teau, Y. Matsuo and M. Umeki }

\ \ 

\centerline{Department of Physics, University of Tokyo,}
\centerline{7-3-1 Hongo, Bunkyo-ku, Tokyo 113, Japan} 

\ \ 

\centerline{ (Received, \hskip 5cm \ \ ) }

\ \ 

\begin{abstract}

A generalized theory of two-dimensional isotropic turbulence is developed 
based on conformal symmetry. A number of minimal models of conformal 
turbulence are solved under an extended constraint including both the 
enstrophy cascade by Kraichnan and the discontinuity of vorticity by 
Saffman. There are an infinite number of solutions which fall into two 
different categories. An explicit relation is derived in one of the 
categories between the central charge of Virasoro algebra, the lowest 
anomalous dimension and the power of the energy spectrum. Some statistical 
properties such as energy spectrum, skewness, flatness and Casimir 
invariants are predicted and compared with numerical simulation by the 
pseudospectral method. 

\end{abstract}


\ \ 

\section{Introduction}

The problem of the energy spectrum has still been controversial 
in two-dimensional turbulence. 
In three dimensions, dimensional analysis by Kolmogorov \cite{Ko} 
shows that 
the energy spectrum in the inertial subrange should be scaled as 
$E(k)= C\ep^{2/3}k^{-5/3}$, where $\ep$ is the energy dissipation rate. 
Kraichnan \cite{Kr}, Leith \cite{Le} and Batchelor \cite{Ba} applied 
Kolmogorov theory to the two-dimensional turbulence and 
obtained that a power of the energy spectrum should be $-3$ 
according to existence of the enstrophy cascade in the 
inviscid limit. 
On the other hand, Saffman \cite{Saf} derived the power $-4$ 
of the energy spectrum by assuming the jump of constant-
vorticity layers. 
A possible explanation of these difference 
is proposed by Gilbert \cite{Gi} in which spiral structures of vortical 
region play an important role and may fill the gap 
between these two spectra. 

In many of numerical simulations \cite{Ki}\cite{Br}\cite{San} 
of decaying two-dimensional turbulence with the initial gaussian 
spectrum, the power increases from $-4$ to $-3$, contours of 
vorticity are stretched and an active enstrophy cascade is observed 
up to $t\approx 10$. 
Then coherent vortices emerge with spiral tails 
and the power of the energy spectrum decreases \cite{Mc}.
Moreover, a very long-time simulation \cite{Mo2} shows 
that in the final stage the flow seems to settle down to 
the universal "sinh" state with the maximum entropy, 
which is predicted by a statistical 
argument of the assembly of point vortices \cite{Mo1}.
Analytical closure theories, such as direct interaction 
approximation (DIA) and test field model have been also applied to 
evaluate quantitatively the energy spectrum with its constant and 
the enstrophy transfer \cite{Kr2} \cite{He} in a reasonable 
comparison with numerical results. 
Kraichnan \cite{Kr2} argues that the 
logarithmic correction can escape from the divergence of the 
enstrophy integral. 

Recently, Polyakov \cite{P1} \cite{P2} has proposed that
one may use the conformal field theory (CFT) 
to analyze two dimensional turbulence. 
Since there appears power law in the correlation functions
of velocity fields, for example,
the existence of  the conformal invariance seems quite natural.
The merit to use CFT is that there are well-classified models
(minimal models)
where there appears only finite number of independent operators.
Usually, when we try  to calculate $n$ point functions, we need
the knowledge of $n+1$ point functions because of the nonlinearity of
the Navier-Stokes equation.  Since we can not usually cope with 
infinite number of degree of freedom, we need to restrict them to
the physically relevant ones from outside, for example, quasi-normal 
and DIA theories. 
If the turbulence is described by CFT minimal models, such
restriction comes from mathematical consistency and there are
no arbitrariness.
In the original paper, Polyakov proposed a minimal model  
(21,2) as the candidate which may describe turbulence.
This model predicts quite a reasonable value for 
the exponent.
However, it was indicated later by  
several groups \cite{Ma}\cite{Lo}\cite{FH1}\cite{FY}
that Polyakov's condition on the CFT  is not 
satisfied by unique model, but in fact 
there are infinite number of candidates. 
Also, as Polyakov himself discovered \cite{P2}, 
there may be some ambiguities in the
conditions 
for CFT if one considers the non-vanishing of one point functions
in the system with boundaries.
Even if one takes the modified conditions, the situation  does not
change and there are again a lot of CFT candidates.\cite{CNPS2}
In any case, if one can not find other appropriate  conditions, 
one may not tell which model is more preferable. One can not even say 
whether there are several universality classes. 

Due to appearance of infinite solutions, we can not predict uniquely
the exponent of energy spectrum.  
In such situation, it is difficult to 
convince ourselves whether prediction from CFT is useful
as long as we consider only one observable.
One way to proceed in this situation is to consider the global
(or geometrical) aspect of the conformal turbulence.\cite{CNPS1}
In this paper, we would like consider another direction.
In CFT models, there appear a lot of primary fields
beside the original velocity field.  It is quite important
consider
the physical meaning of these fields.
Some of them can be obtained in the short distance
expansion of the velocity field.
Thus we can relate it to the  the dependence of
three or four point functions on the Reynolds number
which is known as the ``skewness"
or ``flatness".  If the assumption of the local equilibrium is 
correct, they should be some universal constants.
However, in CFT, we get nontrivial dependence on
Reynolds number due to the anomalous dimension.
In three-dimensional turbulence, there is a tendency that 
both skewness and flatness of velocity derivatives 
increases as Reynolds number becomes large \cite{Va}. 
In many of CFT solutions, however, the behavior is the opposite 
to these three-dimensional observations. 

Interestingly, as we prove in this paper,
these exponents are not independent with each other but
can be uniquely determined from the exponent of energy spectrum.
This property is tightly related to the finiteness of minimal models.
Intuitively, in order to calculate the higher
correlation functions, the most dominant contribution comes from
the lowest dimension operator in the OPE.  However, in the minimal models,
there are always lower bound in the dimensions of primary fields.
We surveyed more than 1600 candidates of CFT solutions. In 99\%
of them, the lowest dimension is attained in the first OPE between
$\psi\times\psi$. The other 1\% behaves
quite differently but we show that
they can be determined exactly.

As another peculiar character, 
two-dimensional inviscid flows possess infinite number of 
conserved quantities $<\om^n >$ which are topological invariants 
of area-preserving diffeomorphism, called Casimir. 
It is commonly believed that 
these conservation laws may affect the character of the 
two-dimensional turbulence, i.e. ergordicity, energy spectrum 
and so on, although the effects are not fully understood. 
A finite-mode system of two-dimensional Euler flow 
preserving such conservation laws is derived by Zeitlin\cite{Ze}. 
However, the effect of this conservation laws on the behavior of 
turbulent states is still unresolved. 
Using CFT theories, Falkovich and Hanany\cite{FH2} showed that the 
dependence of the dissipation rate of $<\om^n >$ on the viscosity 
$\nu$ can be predicted. 
We extend their results and compare the prediction with the 
numerical results. 

This paper is organized as follows. In section 2 we review the 
classical theories of two-dimensional turbulence. 
It is shown that the extended theory, which contains both 
the Kraichnan-Leith-Batchelor's and Saffman's theories, 
is useful in order to understand the 
conformal turbulence. In section 3 Polyakov's conformal turbulence is 
reviewed with a brief sketch of conformal field theory. 
The original CFT models are generalized so that the imposed 
constraint includes not only the Kraichnan's constant transfer of 
enstrophy but also the non-cascade theory by Saffman. 
In section 4 solutions of this generalized CFT models of 
two-dimensional turbulence are shown. 
Distributions of the energy spectra are revealed, 
A relation between the central charge in CFT and the power of the 
energy spectrum is described.
In section 5 higher order statistics including skewness and flatness 
factors are derived. 
Scaling of Casimir invariant $<\om^n>$ and its decay rate 
are discussed in section 6. 
These predictions are compared with the numerical simulations 
in section 7. Conclusions and discussions are summarized in section 8. 
Throughout this paper, attention is paid on removal of the gap 
between the field theory and fluid communities.

\bigskip

\section{Theories of two-dimensional turbulence}

We shall review standard theories of two-dimensional turbulence 
in order to understand the conformal theory of it. 
In what follows we mainly keep in mind the decaying turbulence, 
i.e. the system with no external forcing. 
The equations governing the two-dimensional incompressible 
viscous flow are 

\begin{equation}
\frac{\pd {\bv}}{\pd t} + ( {\bv} \cd \nabla ) {\bv} = 
-\nabla p +\nu \De {\bv}, 
\label{t1} \end{equation}
\begin{equation}
\nabla \cd {\bv} = 0.  
\label{t2} \end{equation}
The vorticity equation is given by
\begin{equation}
\frac{\pd \om}{\pd t} + J (\ps, \om) = \nu \De \om, 
\label{t3}
\end{equation}
where $\ps$ is the streamfunction, $\om=-\De \ps$ is the vorticity, 
and the Jacobian $J$ is defined by 
\begin{equation}
J (\ps, \om) = \fr{\pd \ps}{\pd y}\fr{\pd \om}{\pd x} -
\fr{\pd \ps}{\pd x}\fr{\pd \om}{\pd y}. 
\label{t4} \end{equation}
From \Eq{t1} and \Eq{t3}, the evolution of total energy 
$E= \fr{1}{2}<|{\bv}|^2>$ 
and square-vorticity, called enstrophy 
$\Om =\fr{1}{2}<\om^2>$, is given by 
\begin{equation}
\fr{dE}{dt} = -2\nu \Om \equiv -\vep, 
\label{t5} \end{equation}
\begin{equation}
\fr{d \Om}{dt} = -2\nu P \equiv -\et , 
\label{t6} \end{equation}
Because the enstrophy decreases monotonically, 
the energy dissipation rate 
$\vep$ vanishes in the inviscid limit 
$\nu \rightarrow 0$. Thus the stationary energy cascade is 
irrelevant in decaying two-dimensional turbulence, 
although in the forced system, 
the energy is transferred to the low 
wavenumber side, called the inverse cascade. 

In Kraichnan-Leith-Batchelor theory, the turbulence is 
characterized by the cascade of enstrophy. The dimensional 
analysis shows the energy spectrum is expressed as 
\begin{equation}
E(k) = C\eta ^{2/3} k^{-3}, 
\label{t7} \end{equation}
in the inertial subrange $ L^{-1} \ll k \ll \de^{-1} $
where $\eta$ is the enstrophy dissipation rate, 
$C$ is a universal constant, 
$L$ denotes the large scale length which may be considered as 
the size of the largest eddy of coherent structures or the 
boundary condition, and $\de$ implies the dissipation scale 
given by
\begin{equation}
\de = \nu^{1/2}\eta^{-1/6}. 
\label{t8} \end{equation}
The total energy estimated by the integral in the inertial range 
leads to 
\begin{equation}
E= \int_{L^{-1}}^{\de^{-1}} E(k) dk = \fr{1}{2}C\eta^{2/3} L^2.  
\label{t9} \end{equation}
On the other hand $L$ may be evaluated by $E$ and $\Om$ as 
\begin{equation}
L = \Bl( \fr{E}{\Om} \Br)^{1/2}. 
\label{t10} \end{equation}
From \Eq{t9} and \Eq{t10} we may obtain the Reynolds number as 
\begin{equation}
Re \equiv \fr{E}{\nu \sqrt{\Om}} \approx \Bl(\fr{L}{\de}\Br)^2 . 
\label{t11} \end{equation}
In this theory $\eta$ and $\nu$ are considered as independent 
and it is allowed that $\eta$ remains finite in the inviscid limit. 

Saffman considered that, in the context of decaying 
two-dimensional turbulence, the discontinuity of the vorticity is 
the asymptotic feature by extending the argument of one-dimensional 
turbulence, i.e. Burgers equation. 
The energy spectrum may be given by 
\begin{equation}
E(k)= \fr{\Om}{2L} k^{-4}, \ \ \ \ L^{-1} \ll k \ll \de^{-1}, 
\label{t12} \end{equation}
and the dissipation length is given by 
\begin{equation}
\de = \fr{(\nu L)^{1/3} }{ \Om^{1/6} }. 
\label{t13} \end{equation}
The enstrophy dissipation rate is estimated as 
\begin{equation}
\et = \fr{\nu^{2/3}\Om^{7/6} }{ L^{4/3} }.  
\label{t14} \end{equation}
Using $L^2 \Om = E $ and $E$ remains constant in the inviscid limit, 
\Eq{t14} is rewritten as 
\begin{equation}
\et = \fr{\nu^{2/3}\Om^{11/6} }{ E^{2/3} }.  
\label{t15} \end{equation}
Thus $\eta$ depends on $\nu$ as $\nu^{2/3}$. 
By Saffman's picture there is no enstrophy cascade 
in the inviscid limit. 
Reynolds number is given by 
\begin{equation}
Re = \Bl( \fr{L}{\de}\Br)^3. 
\label{t16} \end{equation}

Comparing the energy spectra and the dissipation lengths 
of the two theories, it seems quite useful to develop a following 
extended theory. That is, the energy spectrum in the inertial range 
is given by 
\begin{equation}
E(k) = \fr{\Om}{2} L^{\al +3}k^{\al },
 \ \ \ L^{-1} \ll k \ll \de^{-1},    
\label{t17} \end{equation}
and the dissipation scale is expressed as 
\begin{equation}
\de = L \Bl( \fr{\nu  }{ \sqrt{\Om} L^2 } \Br)^{\be}, 
\label{t18} \end{equation}
where parameters $\al$ and $\be$ are introduced in order to bridge 
over the two standard theories. 
This generalization is also important when we consider 
CFT turbulence described in the following section.

From \Eq{t18}, the Reynolds number is estimated as 
\begin{equation}
Re = \Bl( \fr{L}{\de}\Br)^{1/\be}. 
\label{t19} \end{equation}
The enstrophy dissipation rate $\eta$ is estimated as 
\begin{equation}
\eta = 2\nu \int_{L^{-1}}^{\de^{-1}}k^4 E(k) dk 
\approx \fr{\nu \Om L^{\al +3} }{\de^{\al +5}} . 
\label{t20} \end{equation}
Substituting \Eq{t18} into \Eq{t20} the dependence of $\eta$ on 
$\Om , \nu , L $ is expressed as 
\begin{equation}
\eta \approx \Om^{1+\ga/2}  
\nu^{1-\ga} L^{-2 + 2\ga },
\label{t21} \end{equation}
where $\ga= \be(\al+5)$. 
Using \Eq{t10} $\eta$ is expressed by $\Om, \nu, $ and $E$ as 
\begin{equation}
\eta \approx \fr{\Om^2 \nu }{ E} \Bl(\fr{E}{\nu \sqrt{\Om}}
\Br)^{\ga}.
\label{t22} \end{equation}
Moreover, the restrictions 
\begin{equation}
(\al, \be) = (-3, \fr{1}{2}),  \ \ \  (\ga =1), \ \ \ \ 
(\al, \be) = (-4, \fr{1}{3}),  \ \ \  (\ga =\fr{1}{3}), \ \ \ \ 
\label{t23} \end{equation}
may be imposed if this extension should meet both the enstrophy 
cascade theory and Saffman's spectrum. 

A more physical argument, which reconcile the two theories 
similarly to our idea, is made by Gilbert \cite{Gi}, 
who showed that the energy spectra with powers 
between $-3$ and $-4$ can be derived by considering that the 
isolated strong vortex cores wind up the discontinuous 
lines of the vorticity. This spiral model gives the 
energy spectrum in the inertial subrange as 
\begin{equation}
E(k) \approx (\Gamma t)^{2/(s+1)}k^{-4+(s-1)/(s+1)}, 
\label{t24} \end{equation}
where $\Gamma $ denotes the strength of the vortex and 
the parameter $s$ specifies the distribution of 
the vorticity in the core.  The value of $s$ is 
allowed as $1<s<\infty$, $s=1$ implies the Saffman's power $-4$, 
$s=\infty$ gives the KLB values $-3$ and the index of the 
realistic vortex is approximately 2 
( the value of the point vortex ), which gives the power 
$-11/3$. 

In the case of decaying turbulence, the temporal behavior of $\Om$ 
can be obtained by integrating \Eq{t22} 
assuming that $E$ is nearly constant: 
\begin{equation}
\Om(t) = \Bl[ \Om(0)^{-1+ \ga/2} + \Bl(1-\fr{\ga}{2} \Br)
\Bl( \fr{\nu}{E}\Br)^{1-\ga} t \Br]^{\fr{2}{\ga-2 }}.  
\label{t25} \end{equation}
Thus the enstrophy decays at large $t$ as 
\begin{equation}
\Om\approx 
\Bl(\fr{\nu}{E} \Br)^{\fr{2(1-\ga)}{\ga-2} }
t^{\fr{2}{\ga-2} },  
\label{t26} \end{equation}
and the enstrophy dissipation rate becomes 
\begin{equation}
\et \approx 
\Bl(\fr{\nu}{E} \Br)^{\fr{\ga (1-\ga)  }{\ga-2 } }
t^{\fr{4-\ga }{\ga-2 }}. 
\label{t27} \end{equation}
These are consistent with $\Om \sim t^{-2}, \et \sim t^{-3}$ of 
Batchelor if $\ga=1$ 
and $\Om \sim t^{-6/5}, \et \sim t^{-11/5}$ of Saffman if $\ga= 1/3 $. 

We are also interested in the behavior of the infinite number of 
inviscid invariants, i.e., arbitrary powers of vorticity, in the limit 
$\nu \ra 0$. 
The decay rate of $\Om_n \equiv \fr{1}{2} <\om^n>$ is given by 
\begin{equation}
\et_n \equiv -\fr{d\Om_n}{dt} = 
\fr{n(n-1)\nu}{2} < \om^{n-2}( \nabla \om)^2> . 
\label{t28} \end{equation}

It is difficult to get the dependence of $\et_n$ on $\Om$ and $\et$, 
but if we assume a random phase distribution of the Fourier 
coefficients of $\om$, we get the dependence as 
\begin{equation}
\et_{n}= 2^{\fr{n}{2}-1}n (n-1)(n-3)!! \nu 
\Om^{\fr{n}{2}-1} P \propto \Om^{\fr{n}{2}-1} \et,
\label{t29} \end{equation}
for an even $n$ and $\et_n = \Om_n =0$ for an odd $n$. 
From \Eq{t26},\Eq{t27} and \Eq{t29}, $\eta_{n}$ decays as
\begin{equation}
\et_{n} \propto 
\Bl(\fr{\nu}{E} \Br)^{ \fr{(1-\ga) [n-2+ \ga] } 
{ \ga-2 } } t^{\fr{n+2-\ga}{ \ga-2 } }.
\label{t30} \end{equation}
The pair $(\al, \be )=(-3,1/2)$ gives 
$$
\et_{n} \sim t^{-n-1}, 
$$
and $(\al, \be )=(-4,1/3)$ leads to 
$$
\et_{n} \sim (E /\nu)^{(6n-10)/15} t^{-1-3n/5}.
$$ 
Thus the dissipation rate of $\Om_{2n}$ 
is independent of viscosity by the enstrophy cascade theory 
and increases as $\nu$ tends to zero by Saffman's picture  
in the random phase approximation. 

\section{Conformal turbulence}

\subsection{Short summary of CFT}

In order to make smooth communication between fluid dynamics
and elementary particle community, we would like
give a short sketch of some  basics of CFT. 
The material
is quite standard. See the original paper 
to understand the detail\cite{BPZ}.
Also there are many nice review
articles, for example,
see \cite{ID}. 

In quantum field theory, symmetry generally plays an important role.  
On the one hand symmetry implies existence of conservation law and resulting
 in effective 
reduction of the degrees of freedom. 
On the other hand, symmetry is a useful tool to select out 
nicely behaving class of  field theories among vast amount of all 
the possible field theories.
 
The most successful example of such is the class of two dimensional conformal 
field theories (CFT).  The field theory in the class is defined in 
two dimension, and  has the conformal invariance as its symmetry.   
This means, if we use a complex variable  
$z=x_1+ix_2$, dynamics of theory is invariant under  an complex analytic 
transformation $z'=f(z)$.   
Polyakov et al. \cite{BPZ} performed a systematic study of CFT and 
showed that the conformal symmetry is really powerful to classify possible
 CFTs and solve them.

Physically, CFT concerns 
the critical point of the second order phase transition of
statistical systems.  On the critical point, 
we are faced with the situation that the correlation 
length of the operators in the system grows up to infinity.
In this limit, the system gains the conformal invariance. 
Then the two dimensional system is described by  CFTs.  

Especially in two dimensions, it is known that 
the conformal symmetry becomes infinite dimensional
and generated by the operators 
$L_n$ and $\Lb_n$ $(n=0,\pm 1,\pm 2,\cdots)$.   They 
form a Lie algebra:
\begin{equation}
	\cmmt{L_n,L_m}=(n-m)L_{n+m}+\frac{c}{12}(n^3-n)\delta_{n+m,0},
	\label{virasoro}
\end{equation}
which is called Virasoro algebra.  The last term in the RHS comes from 
 quantum anomaly. 

Important fact is that we have  a detailed understanding of 
its representations theory, 
and he knowledge is really helpful in classifying all the possible CFTs.  
Representation theory is quite different whether $c\geq 1$ or $c< 1$. 
In the former case, the classification of the possible theories 
is not completely done with Virasoro algebra alone, 
we need further informations such as
the extended symmetry to classify and solve the system.
 Conversely, in the latter case,
the Virasoro algebra is strong enough to completely classify 
all the possible theories and to solve the system
completely.  In fact, it was shown that 
the central charge is restricted
to take only the discrete values,
\begin{equation}
	c=1-\frac{6(p-q)^2}{pq},
	\label{c5}
\end{equation}
with $(p,q)$ mutually coprime integers. These models are
so-called minimal models.

In CFT, we have two kinds of fields called primary fields and 
secondary fields. 
The former is covariant under  the conformal transform,
\begin{equation}
	\phi_{\Delta\Deltab}(z,\zb)\rightarrow
	\left(\frac{\partial w}{\partial z}\right)^\Delta
	\left(\frac{\partial \wb}{\partial \zb}\right)^{\Deltab}
	\phi_{\Delta\Deltab}(w,\wb), 
	\label{c6}
\end{equation}
while the latter is not.   
For each primary field, there are infinite number of
secondary fields. Those can be obtained from the primary field
through the action of the Virasoro generators.
Derivative of the primary field is a typical secondary field. 

One of the most important properties of the minimal model
is that the number of primary fields is finite.
For the minimal model characterized by $p,q$,
the conformal dimensions $\Delta$ ($\Deltab$) of primary fields
are restricted to take the following discrete values,
\begin{equation}
	\Delta_{rs}\equiv 
	\frac{(ps-qr)^2-(p-q)^2}{4pq}\qquad  1\leq r\leq p-1,
	\quad 1\leq s\leq q-1,
        \quad ps > qr.
	\label{c7}
\end{equation}
The final restriction $ps > qr$ comes from the fact that we can
identify primary field with label $(r,s)$ and $(p-r,q-s)$.
This set of primary fields is called the Kac table.

It may be intuitive to compare the representation of
the Virasoro algebra with that of $su(2)$ algebra.\footnote{
There are no object in $su(2)$ algebra which correspond
to the central charge of the Virasoro algebra.
This term originates from the infinite dimensionality
of the algebra.}
The primary field described above corresponds to the
highest weight state in the $su(2)$ representation.
The states generated from highest weight state 
by the action of lowering operator
should be compared to the secondary field.  
Finally, corresponding to the composition rule of the angular momenta, 
two primary fields are composed to give 
several other primary fields.  Namely the primary fields form an algebra 
by the composition operation.   
The composition rule is called  fusion algebra or operator product expansion, 
OPE.  

The rule can be deduced from
the representation theory just like the computation of the
Clebsh-Gordan coefficients of $su(2)$ representation.
For the Virasoro algebra, the OPE is explicitly given by,
\begin{equation}
	\phi_{r's'}\times\phi_{r''s''}
	\sim \sum_{r=|r'-r''|+1}^{min(r'+r''-1,2p-r'-r''-1)}
	\sum_{s=|s'-s''|+1}^{min(s'+s''-1,2q-s'-s''-1)}
	\cmmt{\phi_{rs}},
	\label{fusion}
\end{equation}
where $\phi_{r,s}$ denotes the primary field whose conformal dimension is 
$\Delta_{rs}$. The symbol $\cmmt{\quad }$ denotes the conformal family
(the primary field and its associated secondary fields).
If we regard \Eq{fusion} as the short distance expansion,
the contribution from the secondary fields are higher order
in power of $z\zb$ compared to those coming from the primary field, so 
they are irrelevant.

Finally, let us see some examples of CFT.  
The minimal models are not necessarily unitary. In 
other words, negative norm state appears in some situations.
Only if  $p=q+1=m+1$ with $m=2,3,4...$, the theory is unitary.
Many of the critical points of the known 2-dim statistical systems 
are described by the models in this series. 
For example, the Ising model ($c=1/2$, i.e. $m=3$)
the tri-critical Ising model ($c=7/10$, i.e. $m=4$),
the three state Potts model ($c=4/5$, i.e. $m=5$),
and so on.  For these models, one can find the physical interpretation
for the primary fields.  For example, for Ising model, 
we have three independent primary operators,
$$
	\Delta_{11}=0\quad
	\Delta_{21}=1/16 \quad
	\Delta_{31}=1/2.
$$
These operators may be identified with, identity operator,
energy density and the spin field respectively. 

On the other hand, the nonunitary models are by 
no means  physical, but contains some very interesting models,
such as Yang-Lee edge singularity ($c=-22/5$, i.e. $(p,q)=(5,2)$).
They are characterized by the appearance of the negative dimension 
operators.  For the Yang-Lee edge singularity case,
we have $\Delta_{2,1}=-1/5$.  In the unitary case,
the lowest dimension operator in the system is always the vacuum,
which is represented by the primary field with $\Delta_{1,1}=0$.
For the non-unitary case, this 'vacuum' is no longer the lowest
dimension operator but rather we have a primary field with
dimension,
\begin{equation}
	\Delta_0=\frac{1-(p-q)^2}{4pq}.
	\label{lowestdim}
\end{equation}
This operator may be regarded as the 'true vacuum' of the system.

Let us take a nonunitary model $(p,q)=(21,2)$ as an example.  
Field content consists of ten fields:
$\phi_{1,1},\phi_{2,1},\cdots,\phi_{10,1}$.  Since the second index $s$ 
is always unity in this case, we will omit it hereafter.  
OPE is, for example 
\begin{equation}
	\phi_4\times\phi_4
      \sim \cmmt{\phi_7}+\cmmt{\phi_5}+\cmmt{\phi_3}+\cmmt{\phi_1}. 
 \end{equation}
For the two point function, this fusion
rule implies that 
\begin{eqnarray}
   <\phi_4(z)\phi_4(0)> &=& 
   (z\zb)^{2\Delta_7-4\Delta_4}<\phi_7(0)>+ \cdots +
  (z\zb)^{2\Delta_1-4\Delta_4}<\phi_1(0)>
  \non\\
  &&+ \ \mbox{higher order terms w.r.t. $z\zb$}.
 \end{eqnarray}
Note that $z,\zb$ dependence of each term in RHS is determined by 
the accordance of the length dimension in both sides.

\subsection{Conformal Turbulence}
\subsubsection{Polyakov's original definition}

In CFT approach to two dimensional turbulence, the fluid circulation 
function $\psi$ is supposed to be described by
a primary field of a minimal model of CFT. 
$\psi$ is related to the velocity field through 
$v_\alpha=\epsilon_{\alpha\beta}\partial_\beta \psi$.
Polyakov's conditions for CFT to describe turbulence
are following,
\begin{enumerate}
	\item  Hopf equation should be satisfied,
	\begin{equation}
		<\epsilon_{\alpha\beta}\partial_\alpha\psi\partial^2
		\partial_\beta
		\psi\cdots>=0.
		\label{1}
	\end{equation}
	\item  Constant flow of enstrophy in the momentum.
\end{enumerate}
Assuming operator product expansion (OPE), 
\begin{equation}
	\sq{\psi}\sq{\psi}\sim \sq{\phi}+\cdots,
	\label{2}
\end{equation}
where $\phi$ is the smallest dimension primary field in OPE,
the nonlinear term in the Navier-Stokes equation
\Eq{1} is equivalent to
\begin{equation}
	|a|^{2\Del{\phi}-4\Del{\psi}}(\Ln{-2}\Lnb{-1}^2-\Lnb{-2}\Ln{-1}^2)\phi,
	\label{3}
\end{equation}
where $a$ is a viscous cutoff length. 
Vanishing of this quantity then produces either
\begin{equation}
	2\Del{\phi}-4\Del{\psi}>0,
	\label{3.5}
\end{equation}
or
$
	\Ln{-2}\phi \sim \Ln{-1}^2 \phi.
$
The second equation (which is equivalent to
that $\phi$ is (2,1) type primary field) 
is not compatible with the conditions
on enstrophy and Polyakov has picked up the first
condition.

The constant enstrophy flow constraint is more subtle.
The time derivative of the enstrophy gives
\begin{equation}
	\Pa{\Omega}{t}=<\omega\phi^{(2)}>,
	\label{4}
\end{equation}
with $\omega =\partial^2 \psi$ and
$\phi^{(2)}=(\Ln{-2}\Lnb{-1}^2-\Lnb{-2}\Ln{-1}^2)\phi$.
Rewriting the right hand side of \Eq{4} as the integral
in the momentum space, the constant enstrophy condition 
is equivalent to
\begin{equation}
	<\omega(x)\phi^{(2)}(0)>\sim x^{0}.
	\label{5}
\end{equation}
Naive dimensional counting shows that it is equivalent to
$\Del{\psi}+\Del{\phi}+3=0$. However, we have to be careful to
the contribution of the one point function, which is non-vanishing
in the general topology of two dimensional space.
The dependence of one point function on the characteristic length
of the system is
\begin{equation}
	<\psi>=L^{-2\Del{\psi}},
	\label{6}
\end{equation}
for any primary field $\psi$.
For unitary theory where dimension of any primary field
is positive, one point function is not relevant in the 
large $L$ limit.  
However, since we are discussing on the system where
dissipation takes place, we are dealing with non-unitary theory 
where negative dimensional operator appears.  In this
situation, the lowest dimension operator becomes most relevant one.
If we denote $\chi$ as the lowest dimension primary field
in OPE $\sq{\psi}\sq{\phi}$, then the correct condition
for the constant enstrophy flow is
\begin{equation}
	\Del{\psi}+\Del{\phi}-\Del{\chi}+3=0.
	\label{7}
\end{equation}

The solutions of CFT that satisfy \Eq{3.5} and \Eq{7} was discussed by
Chung et. al.\cite{CNPS2}  and 
they found several hundreds candidates.

Once we know that there are solutions, the next step is how 
to extract physical exponents out of them.
The most important one is discussed by Polyakov, i.e.
the exponent of energy spectrum,
which may be described by,
\begin{equation}
	E=\int dk E(k)=\frac{1}{2}\sum_{\mu=1}^2<v^\mu v^\mu>.
	\label{8}
\end{equation}
From contribution of  the one-point function \Eq{6},
one finds that
\begin{equation}
	E(k)\sim k^\alpha,\qquad \mbox{with}\quad 
	\alpha={4\Del{\psi}-2\Del{\phi}+1}.
	\label{9}
\end{equation}
If the one-point correlation integral vanishes, 
the energy spectrum becomes
\begin{equation}
	E(k)\sim k^\alpha,\qquad \mbox{with}\quad \alpha={4\Del{\psi}+1}.
	\label{9d}
\end{equation}
These expressions are Fourier transform of  two-point functions,
\begin{equation}
	<\psi(z)\psi(0)>\sim |z|^{-4\Delta_\psi},
\end{equation}
for correlation function on sphere \cite{BPZ}
and 
\begin{equation}
	<\psi(z)\psi(0)>\sim L^{-2\Delta_\phi}
	|z|^{-4\Delta_\psi+2\Delta_\phi},
	\label{torus}
\end{equation}
for correlation function on general topology Riemann surface
including torus.
This kind of correlation functions was discussed previously 
by J. Cardy \cite{C} which is the refinement of the discussion
given by Fisher \cite{F}. The correlation function of
unitary model (in particular, Ising model) on the torus
was obtained explicitly by Di Francesco et. al.\cite{FSZ}.
In the turbulence context, some of
the physical implication was  discussed by Polyakov\cite{P2}
and Chung et.al. \cite{CNPS1} and \cite{CNPS2}.
The most important observation is that we can not neglect
the contribution of the one point function.  This is because
the dimension of $\phi$ is negative and it becomes dominant
in the large $L$ limit in \Eq{torus}.

\subsubsection{Generalized conformal turbulence}

Let us introduce the other possibilities that can be
imposed on the conformal field theory.
As we saw in section 2, there are many variants of the
two-dimensional turbulence models where the enstrophy
flow may depends on the viscosity.
Also, the condition will be changed if one have vanishing
one point function which was originally considered by Polyakov.
To get the corresponding object in the conformal turbulence,
we need to compare the dissipation calculated by
the viscosity term  and
nonlinear term in the Navier-Stokes equation.
The correlation functions which we need to use is,
\begin{eqnarray}
	<\omega(z)\omega(0)> & = & L^{-2\Delta_\phi}
	|z|^{-4\Delta\psi-4+2\Delta\phi}.
	\non \\
	<\omega(z)\delta_2\phi(0)> & = & 
	L^{-2\Delta_\chi}|z|^{-2\Delta_\psi-2\Delta_\phi+2\Delta_\chi-6}.
	\label{gc2}
\end{eqnarray}
If we denote the short distance cutoff by $\delta$,
the dissipation rate of the enstrophy can be either computed as,
\begin{equation}
	-\frac{d\Omega}{dt}
	=\eta\approx\nu\delta^{-6-4\Delta_\psi+2\Delta_\phi},
	\label{gc3}
\end{equation}
by using the viscosity term, or can be alternatively given by,
\begin{equation}
	-\frac{d\Omega}{dt}=\eta\approx
	\delta^{-6-2\Delta_\psi-2\Delta_\phi+2\Delta_\chi},
	\label{gc4}
\end{equation}
by using the nonlinear term.
In order to balance those two, we have to impose,
\begin{equation}
	\nu\approx \delta^{2\Delta_\psi-4\Delta_\phi+2\Delta_\chi}.
	\label{gc5}
\end{equation}
The similar expression when we have vanishing one-point function
is discussed by Hanany et. al.,
\begin{equation}
	\nu \approx \delta^{2\Delta_\psi-2\Delta_\phi}.
	\label{gc6}
\end{equation}
By employing these relations, one may obtain the 
dependence of $\eta$ on the viscosity as
\begin{equation}
	\eta\approx \nu^{1-\gamma},\quad \gamma=\gamma_1=
	\frac{2\Del\psi-\Del\phi+3}{
	\Delta_\psi-2\Delta_\phi+\Delta_\chi}. 
	\label{gc7}
\end{equation}
If we use the vanishing one-point function condition,
$\gamma_1$ should be replaced by,
\begin{equation}
	\gamma=\gamma_2=\frac{2\Del\psi+3}{\Delta_\psi-\Delta_\phi}.
	\label{gc8}
\end{equation}
In either case,  if we employ the Kraichnan-Leith-Batchelor type 
condition, one has to impose,
\begin{equation}
	\gamma_i=1,\qquad i=1,2,
	\label{gc9}
\end{equation}
and if we use Saffman type condition, we get,
\begin{equation}
	\gamma_i=\frac{1}{3}, \qquad i=1,2.
	\label{gc10}
\end{equation}

On the other hand, the parameter $\beta$ 
defined in \Eq{t19}   
may be expressed by the conformal dimensions as
\begin{eqnarray}
	\beta_i & = & \frac{\gamma_i}{\alpha_i+5},
	\non \\
	\beta_1 & = & (2\Del\psi-4\Del\phi+2\Del\chi)^{-1},
	\non \\
	\beta_2 & = & (2\Del\psi-2\Del\phi)^{-1},
	\label{gc11}
\end{eqnarray}
by comparing with (\ref{gc5}-\ref{gc6}).
In this way, one can relate the dimensions of CFT primary fields
with all the parameters which appeared in the previous section.

\section{CFT models}

In order to compute concrete values for the critical exponents, we
need to solve (generalized) Polyakov's conditions \Eq{gc7} 
or \Eq{gc8} explicitly.
This program is rather hard to be carried out
 analytically, hence, we have
only solutions obtained from numerical computation.
Previously, some of the CFT models of
KLB type was found by
\cite{Ma} for vanishing one-point function case and by
\cite{CNPS2} for non-vanishing one point function case.
In those papers, although CFT models are derived, the characteristic
features which are common for those models are not discussed.
In this paper, we study more extensive study for
the generalized conditions \Eq{gc7} 
or \Eq{gc8} to get insights on the generic features
of models which meet these conditions.
We surveyed generalized
Diofantian equations \Eq{gc9} \Eq{gc10}
(we shall call them as P$i$, and S$i$ (with $i=1,2$)
in the following) up to $p<1000$ and $q<100$.
As expected we got many CFT candidates, 1658 models for condition P1,
78 models for condition P2, 67 models for condition S1 and 75 models
for condition S2.
Later we prove that there are infinite number of solutions for 
any conditions of type \Eq{gc7}.


Since we have several CFT models, we can not determine the
exponents in the two dimensional turbulence uniquely.
In figure 1a-d, we show the histogram
of number of CFT models against the exponent of energy spectrum
or each of the conditions.
Where as in figure 1b,1d, we have rather flat spectrum
in the range $-5< \alpha <-3$, we have broader 
($ -5< \alpha <1 $) and localized spectrum 
(peak located around $\alpha = -4.5, \ -3.5, \ -2.5, \ -1.5,
\ -0.5, \ 0.5$) for the modified condition. We can also notice
that a significant portion (82\%) of the CFT models has their
exponent in the range $-5<\alpha<-4$.

We remark that in the course of computation, we need to
determine operator $\phi$ as the lowest dimension operator
in the OPE $\psi\times\psi$.  By this process, 
we can classify the obtained models into two groups.

The first group, which we call category A below, 
is characterized by the fact that $\phi$ has already 
the dimension very close to the lowest dimension $\Del{0}$ of the
system.  In this case, the dimension of $\chi$ is very close
to $\phi$ since both of them are roughly same as $\Del{0}$.
The situation remains same even if we take higher OPEs.
Most of the solutions which we obtained belong to this category.

On the other hand, in the second class of solutions (category B),
neither $\phi$ nor $\chi$ have dimension
closer to  $\Del{0}$. In this case, in every step we take OPE,
we meet lower dimensional operator.
Since Category B is simpler and can be classified completely,
we start with it first.

\subsection{Category B}
The most interesting feature of
solutions of this class is that we can find
an infinite number of solutions for the condition of type
\Eq{c7} for {\em any fractional} $\gamma$.
As far as we have observed, the solutions of
those type always take the following form, i.e.
\begin{eqnarray}
	(p,q)&=&(p_1(n-1)^2, q),\non\\
	\psi&=&(n,1),\non\\
	\phi&=&(2n-1,1),\non\\
	\chi&=&(3n-2,1).
	\label{B1}
\end{eqnarray}
It is obvious from \Eq{fusion} that there appears only
primary fields of type $(r,1)$ in the OPE of primary fields
with index $s=1$.
On the other hand, the conformal dimension of fields $(r,1)$ decreases as
$r$ increases as long as $r <p_1/q(n-1)^2$.
One may confirm that $\phi$ is the minimum dimension in OPE
$\psi\times\psi$ and so is $\chi$ in OPE $\psi\times\phi$ if
\begin{equation}
	n>\frac{3y+2+\sqrt{9y^2+4y}}{2},\qquad y=q/p_1.
	\label{BND}
\end{equation}
Also, for sufficiently large $n$, one may easily convince oneself
that neither $\phi$ nor $\chi$ have dimension closer to
the minimum dimension of the system.

Now if we put those combinations into \Eq{c7},
we get,
\begin{equation}
	\eta\sim\nu^{1-\gamma}\qquad
	\gamma\equiv \frac{2\Del\psi-\Del\phi+3}{
	\Delta_\psi-2\Delta_\phi+\Delta_\chi} =\frac{6p_1-q}{q}.
	\label{B2}
\end{equation}
The right hand side does not depend on $n$.
For any fractional value for $\gamma$, we can adjust $p_1$ and
$q$ such that \Eq{B2} can be satisfied.  For example, to achieve
KLB type condition $(\gamma=1)$, one may put
\begin{equation}
	p_1=1,\qquad q=3.
	\label{B3}
\end{equation}
For Saffman type condition $(\gamma=1/3)$, one may get,
\begin{equation}
	p_1=2,\qquad q=9.
	\label{B4}
\end{equation}
These give the solutions to the Diofantian equation
for $n>10$ in KLB case and for $n>14$ for Saffman case.
This simple observation shows that there are infinite number of
solutions for the condition \Eq{c7} for any $\gamma$.

One of the interesting feature of these solutions is that
all of these models have same energy spectrum for each $\gamma$.
Indeed, one may derive another cancellation of $n$,
\begin{equation}
	\alpha\equiv 4\Del{\psi}-2\Del{\phi}+1=-\frac{q-p_1}{p_1}.
	\label{B5}
\end{equation}
One can remove $p_1$ and $q$ 
from relations \Eq{B2} and \Eq{B5},
\begin{equation}
	\alpha=\frac{\gamma-5}{\gamma+1}.
	\label{B6}
\end{equation}
The parameter $\beta$, on the other hand, 
can be similarly determined by,
\begin{equation}
	\beta=\frac{\gamma}{\alpha+5}=\frac{\ga+1}{6}.
	\label{B7}
\end{equation}

For KLB, one has $(\alpha,\beta)=(-2,1/3)$
 and for Saffman type condition, 
one has
$(\alpha,\beta)=(-7/2,2/9)$. The latter model  seems to fit rather
well to the observed value for the energy spectrum
$-4 < \alpha < -3$.

\subsection{Category A}
Although CFT models in category B
has rather interesting properties, the majority
(99\%) of the solutions we obtained belong to this category.
Although we can not solve the Diofantian equation in general,
we can make several observations for this class.

Interestingly, 
the various exponents of the models in this class
is (although approximately) related to
the  central charge of the system.  The dependence itself
is different for the conditions that we impose on the 
conformal models.  If there are some methods where we can
independently measure the central charge, one may judge
which condition is the appropriate one.  

In order to derive such formulae, it is essential to observe that
\begin{enumerate}
	\item  The operator product expansion of $\psi$ field
	gives rise to a operator $\phi$ whose dimension is very 
	close to the lowest dimension ($\Delta_0$) of the system.
	
	\item  The central charge of the nonunitary
	minimal model is linearly related to $\Delta_0$.
\end{enumerate}

To demonstrate the first fact, we make histograms
of the difference of the conformal dimensions
between field $\phi$ and $\Delta_0$ (figure 2).
We can see that deviation is less than 0.02.  
The second fact can be easily checked by recalling that
$c=1-\frac{6(p-q)^2}{pq}$ and
$\Delta_0=\frac{1-(p-q)^2}{4pq}$.
When $p$ or $q$ is sufficiently large, the first term in 
the denominator of $\Delta_0$ may be neglected and one finds,
\begin{equation}
	\Delta_0=\frac{c-1}{24}.
	\label{c1}
\end{equation}

If we use the generalized condition \Eq{gc7}, we get
\begin{equation}
	\Delta_\psi=\frac{(1-\gamma_1) \Delta_0-3}{2-\gamma_1}.
	\label{c11}
\end{equation}
The energy exponent may be then given by,
\begin{equation}
	E(k)=k^{\alpha_1},\qquad \alpha_1=
	\frac{-2\gamma_1\Delta_0-\gamma_1-10}{2-\gamma_1}.
	\label{c13}
\end{equation}
In particular, 
\begin{equation}
	\alpha_1=-11-2\Delta_0 \quad\mbox{(KLB)},
	\quad -8-\Delta_0 \quad\mbox{(Saffman)}.
	\label{c15}
\end{equation}

Similarly, the viscous scale index $\be$ in \Eq{t18} can be obtained as 
\begin{equation} 
        \be=  \frac{\ga_1-2}{2(\De_0+3) }, 
	\label{c17} 
\end{equation} 
for the condition \Eq{c7}.
From the KLB and Saffman's condition, $\be$ reduces to 
\begin{equation}
	\be=-\frac{1}{2(\De_0+3)} \quad\mbox{(KLB)},
	\quad -\frac{5}{6(\De_0+3)} \quad\mbox{(Saffman)}.
	\label{c19}
\end{equation}

Together with \Eq{c1}, it is clear that 
\Eq{c13} gives the linear relation for the
energy exponent and the central charge.
We plot those relation in figure 3, to show that
these are fulfilled rather nicely.

\subsubsection{Families of infinite solutions in category A}

Although the solutions of category A seem to occur randomly,
there are several models that can be classified
into several families of solutions.
We would like to explain it by using the examples
of condition P1.

First of all, as Chung et. al. has already pointed out, there 
are many occurrences of solutions (in our case, 1347 cases out of 1658),
with special combination of primary fields,
\begin{equation}
	\psi=(26,2), \quad \phi=(15,1),\quad\chi=(30,2).
	\label{21}
\end{equation}
We would like to prove that
this combination becomes the solution of conformal turbulence
whenever
\begin{equation}
	\frac{29}{2} q<p<\frac{31}{2}q,
	\label{22}
\end{equation}
with $p$ and $q$ coprime.  Roughly speaking, this shows that
there are $q$ solutions of this type for each $q$.
This fact explains quite systematic occurrence of
solutions of this type.

The proof is quite elementary.  We must first observe
\Eq{21} solves \Eq{7} for arbitrary $p$ and $q$.
This requirement is equivalent to the algebraic equation
\begin{eqnarray}
	r_1^2+r_2^2-r_3^2-1 & = & 0,
	\non \\
	s_1^2+s_2^2-s_3^2 & = & 0,
	\non \\
	r_1s_1+r_2s_2-r_3s_3-7 & = & 0,
	\label{23}
\end{eqnarray}
if $\psi$, $\phi$, $\chi$ are labeled
$(r_1,s_1)$, $(r_2,s_2)$, and $(r_3,s_3)$.

The next step is to study when $\phi$ described in \Eq{21}
becomes the lowest dimension operator in OPE.
By explicitly comparing the dimension of $\phi$ with
its neighbors, $(13,1)$, $(17,1)$, $(45,3)$, $(45,5)$,
$(47,5)$, one finds that \Eq{22} is the desired condition.
Under this condition, one can easily show that
$\chi=(30,2)$ is indeed the lowest dimension operator
in OPE $\psi\times\phi$. (QED)

For this family of solutions, since the ratio of
$p$ and $q$ is already given \Eq{22}, one can determine
the range of the exponents.  For example,
the range of $\alpha$ is immediately determined by
using $\alpha=3x+\frac{563}{x}-87$,
\begin{equation}
	-4.672< \alpha < -4.177.
	\label{24}
\end{equation}
Since $p$ can take arbitrary (of course, with coprime condition)
integer value in
\Eq{22}, the distribution in this range is quite flat.
This fact is quite consistent with figure 1a.

There are two other families of solutions with similar property,
i.e. they satisfy \Eq{23}.
The first one is,
\begin{equation}
	\psi=(55,4),\quad\phi=(101,7),\quad \chi=(115,8),
	\label{25}
\end{equation}
and the second one is,
\begin{equation}
	\psi=(362,26),\quad\phi=(209,15),\quad\chi=(418,30).
	\label{26}
\end{equation}
There are 54 occurrence of the first family and 2 occurrence of
the second one.  One can similarly show that
they contain infinite solutions as their member. 
The expected range of $\alpha$ is given by
\begin{eqnarray}
	-4.765 & <\alpha< & -4.740,
	\non \\
	-4.998 & <\alpha< & -4.999,
	\label{27}
\end{eqnarray}
respectively.

Although 1418 of 1658 solutions are explainable in this way,
there are still 240 solutions that we can not describe
their origin.
However, we may give some 'phenomenological' reasoning
why there appear peaks around $-3.5$, $-2.5$, 
$-1.5$,$-0.5$, and $0.5$.
As it turns out, those peaks comes from the models
where $\phi$ happen to be $(17,1)$, $(19,1)$, $(21,1)$,
$(23,1)$ and $(25,1)$.\footnote{We remark that the biggest peak
around $-4.$5 comes from models with $\phi=(15,1)$.}
In order that those operators are roughly the lowest
dimension fields, the ratio of $p/q=x$ should be 
17,19,21,23,25 respectively.  Putting these values 
in \Eq{c13}, we get,
\begin{equation}
	\alpha= -3.47,\quad -2.47,\quad -1.48,\quad -0.48,\quad -0.52,
	\label{28}
\end{equation}
respectively. Of these classification, there are 180 models.
%
 
To summarize, although CFT gives infinite number
of solutions for Polyakov's criterion, 
we can extract quite a lot information from them.
Especially, we can determine the relations between
the exponents and have several models which have 
their $\alpha$ in the special range. 

\section{ Higher Order Statistics}

As we saw in the previous sections, the
primary fields $\psi$, $\phi\cdots$ are related to
the observable in two dimensional turbulence through
the energy exponent.
Now some natural question arises. In general, there are
many other primary fields in the minimal models.
How can we extract some informations for those fields?
Here we propose that one direction we may take is to go
further to calculate higher correlation functions.
As long as the assumption of the local equilibrium is correct, if one
makes the dimensionless combination such as,
$<\psi^3>/<\psi^2>^{3/2}$, it should be a
dimensionless constant and there should  be no dependence
on the Reynolds number.  In CFT, this is not true 
because there are singularities in the short distance expansion.

In the correlation function, $<\psi\psi\cdots\psi>$,
the dominant contribution comes from the one point function
of the lowest dimension in the OPE $\sq\psi\times\sq\psi\cdots
\sq\psi$.
This gives the $L$ dependence of the correlation function.
On the other hand,
the singularity coming from the short distance
expansion of $\psi$ fields can be expressed as
power of the short distance cutoff $\delta$.
Let us denote $\phi_n$ as the lowest dimension operator
in the $n$th order product and $\Del{n}$ as the dimension of
$\phi_n$. We may evaluate the $n$ point function in terms of them,
\begin{equation}
	<\psi^n>\sim L^{-2\Del{n}}\delta^{-2n\Del{\psi}+2\Del{n}}.
	\label{sk1}
\end{equation}
It follows that the non-trivial dependence on $\delta$ for the
dimensionless combination,
\begin{equation}
        F_n = 
	\frac{<\psi^n>}{<\psi^2>^{n/2}}\sim (L/\delta)^{-2\Delta_n+n\Del{\phi}}
	\sim Re^{\beta_n}.
	\label{sk2}
\end{equation}
Here $\beta_n$ is the exponent for $n$th order products:
\begin{equation}
	\beta_n\equiv -\beta(2\Delta_n-n\Del{\phi}).
	\label{sk3}
\end{equation}
The simplest choice $F_3$ and $F_4$ corresponds to
what is usually called ``skewness" and ``flatness"
in the fluid dynamics community. 
We shall denote the exponents for $n=3,4$ as $\be_S$ and $\be_F$
in the following discussions.

\subsection{Category A}

For this category, one may substitute $\Del{n}=\Del{0}$
in \Eq{sk3}. Using \Eq{c19}, one obtains
\begin{eqnarray}
	\be_k & = & -\beta(2\Delta_k-k\Delta_\phi)
	\non \\
	 & = & \frac{\Delta_0(k-2)(\gamma-2)}{2(3+\Delta_0)}.
	\label{c21}
\end{eqnarray}

\subsection{Category B}

%
%

Skewness and Flatness may be computed by using,
$\Del{3}=\Del\chi$ and $\Del{4}=\Del{(4n-3,1)}$,
which is valid for $n>2y+1+\sqrt{4y^2+y}$.
Amazingly, even the dependence on parameter $\gamma$ disappears.
\begin{eqnarray}
	\be_S & \equiv &-\beta(2\Del{\chi}-3\Del{\phi})=-\frac{3}{2},
	\non \\
	\be_F & \equiv & -\beta(2\Del{4}-4\Del\phi)=-4.
	\label{15}
\end{eqnarray}
In general, if the minimum dimension operator in $k$th product
of $\psi$ is given by $(kn-k+1,1)$, the $k$th exponent is 
\begin{equation}
	\beta_k=\frac{k(2-k)}{2}.
	\label{16}
\end{equation}
In this way, we find that the solutions for
category B gives {\em unique} prediction for the higher exponents.

\section{Casimir Invariants}

Up to now, CFT is supposed to describe the time independent
aspect of the two dimensional turbulence.
However, the use of Navier-Stokes equation gives
a simple explanation time dependence for the decaying 
turbulence\footnote{
This process seems to be similar to the deformation of
conformal field theory by coupling
the relevant operators to conformal invariant system
discussed by Zamolodchikov\cite{Za}.
In his discussion, the central charge $c$ behaves as the Morse function
in the theory space of CFT and increases its value in the process
of renormalization group. In order to apply his argument to
turbulence, we need to extend his argument to include
the non-unitary theory. 
We would like to 
discuss this aspect in future issues.}.
In particular, we would like to discuss the decay rate
of Casimir invariants $\Om_n$ in detail to compare
the results from CFT to our numerical computation.
One may easily find it as,
\begin{equation}
	\et_n = \fr{n(n-1)\nu}{2} <\om^{n-2}(\nb \om)^2 >
	\propto \nu \de^{2\De_n -2n\De_\ps -2n-2} L^{-2\De_n}. 
	\label{ca1}
\end{equation}
As usual, $\De_n$ is the minimum dimension in the OPE of
$n$ $\psi$ fields.
On the other hand, a dimensional argument shows 
\begin{equation}
	\et_n = \frac{\nu}{\de^2} \Om^{\fr{n}{2}} 
	\Bl( \frac{L}{\de } \Br)^{\al_n}, 
	\label{ca2}
\end{equation}
where $\al_n$ denotes an anomalous dimension for $\et_n$. 
Comparing \Eq{ca1} and \Eq{ca2}, we find 
\begin{equation}
\al_n= 2n(\De_\ps +1) -2\De_n. 
	\label{ca3}
\end{equation}
Using \Eq{t10}, \Eq{t18} and \Eq{ca3}, \Eq{ca2} becomes 
\begin{equation}
\et_n = \Om^{ \fr{n}{2} \{ -\be (\De_\ps +1)+1 \} +1 -\be (\De_n +1 ) }
        \Bl( \frac{E}{\nu} \Br)^{ 2\be(\De_\ps +1 )n -2\be (\De_n -1) -1 }.
	\label{ca4}
\end{equation}
Substituting \Eq{t26} into \Eq{ca4}, we can estimate the decay rate 
of $\Om_n$ as 
\begin{equation}
\et_n \propto \Bl( \frac{\nu}{E} \Br)^{p_n}t^{q_n}, 
	\label{ca5}
\end{equation}
where the powers $p_n$ and $q_n$ can be obtained 
 as
\begin{eqnarray}
p_n &=&  \frac{2\beta(\De_\psi+1)+1-\gamma}{\gamma-2}n
     +   \frac{2\beta(1-\De_n)-\ga}{\gamma-2},
	\non\\
q_n &=& \frac{\beta(-2\De_\psi+1)+1}{\gamma-2} n
        +\frac{-2\beta(1-\De_n)+2}{\gamma-2}. 
	\label{ca6}
\end{eqnarray}

On the other hand, the random phase approximation described in 
section 2 leads to 
\begin{equation}
p_n = (1-\ga)(1+\fr{n}{\ga-2}),
\label{t32} \end{equation}
\begin{equation}
q_n=-1+ \fr{n}{\ga -2}.
\label{t33} \end{equation}
We can see that this difference originates in the anomalous 
dimension. 

\subsection{Category A}

Using \Eq{ca6}, \Eq{c11} and \Eq{c17}, the exponents $p_n$, $q_n$ of 
decay rate can be obtained as 
\begin{eqnarray}
p_n &=& -\frac{2}{\De_0+3}n
        +\fr{2 \{ -\ga_1(\De_0 +1) +\De_0-1 \} }{(\De_0+3)(\ga_1-2)},
	\non\\
q_n &=& -\fr{\De_0+1}{\De_0+3}n
        +\frac{\ga_1(\De_0-1)+8}{(\De_0+3)(\ga_1-2)}. 
	\label{ca7}
\end{eqnarray}
They become
\begin{eqnarray}
p_n &=& -\frac{2(n-2)}{\De_0+3},
	\non\\
q_n &=& -\fr{(\De_0+1)n+\De_0+7}{\De_0+3},
	\label{ca8}
\end{eqnarray}
for the KLB condition and 
\begin{eqnarray}
p_n &=& -\fr{10n+4(\De_0-2)}{\De_0+3},
	\non\\
q_n &=& -\fr{5(\De_0+1)n+\De_0+23}{\De_0+3}, 
	\label{ca9}
\end{eqnarray}
for the Saffman condition. 

\subsection{Category B}

%
%

The decay rate for the higher Casimir invariant can be found from
\Eq{B7}. After some computation, we get
\begin{eqnarray}
	p_n & = & \frac{1}{2(\ga+1)}n^2-
	\frac{8\ga^2+\ga-16}{6(\ga+1)(\ga-2)}n-
	\frac{2(2\ga-1)}{3(\ga-2)},
	\non \\
	q_n & = &
	 -\frac{1}{2(\ga+1)}n^2+\frac{2\ga^2+7\ga-4}{
	 6(\ga+1)(\ga-2)}+\frac{4+\ga}{3(\ga-2)}.
	\label{decayB}
\end{eqnarray}
Unlike category A, these exponents grows as $n^2$.
They become
\begin{eqnarray}
	p_n & = & \frac{1}{4}n^2-\frac{7}{12}n+\frac{2}{3},
	\non \\
	q_n & = & \frac{-1}{4}n^2-\frac{5}{12}n-\frac{5}{3},
	\label{decayB1}
\end{eqnarray}
for the KLB condition and,
\begin{eqnarray}
	p_n & = & \frac{3}{8}n^2-\frac{133}{120}n-\frac{2}{15},
	\non \\
	q_n & = & -\frac{3}{8}n^2+\frac{13}{120}n-\frac{13}{15},
	\label{decayB2}
\end{eqnarray}
for the Saffman condition.

\section{ Numerical Analysis}

In this section, some statistical properties predicted by CFT 
are compared with the numerical data obtained by 
the direct simulation. The pseudospectral method is used, 
which efficiently computes convolutions of the nonlinear terms 
in the physical space and is fully dealiased by the two-third law. 
The modes are 256$\times$256, the time integration is performed 
by the second-order Runge-Kutta scheme up to $t=10$ and the 
width for time step is 0.002. 
We examined five different cases of viscosity, 
$\nu$ = 0.0001, 0.0003, 0.001, 0.003 and 0.01 and main results 
for the first four cases are shown. 
Decaying turbulence created by \Eq{t1} without external forcing 
is investigated. 

Temporal evolution of energy $E$, enstrophy $\Om$ and 
enstrophy dissipation rate $\eta$ is shown in Figures 4a-c.
Initial increase of $\eta$ is observed until $t\approx 5$ for 
$\nu = 0.0001$ and the decay of $\Om$ and $\eta$ 
is prominent after $t=7$. 
Thus, data in the time interval $7 \le t \le 10$ are used to fit 
the asymptotic behavior of $\eta$: 
\begin{equation}
\eta \propto (\nu/E)^p (t+ t_0)^q,            
\label{n1} \end{equation}
where $t$ in \Eq{t27} is replaced by $t+t_0$ in order to get 
a more precise power from the finite-time integration. 
The time offset $t_0$ is estimated by the asymptotic expression of 
$\Om/ \eta \approx a(t+t_0)$, which is independent of $q$ and 
the data $7\le t \le 10$. Numerically obtained values 
of parameters in \Eq{n1} are shown in Table 2. 

Although the numerical power of $t$ varies from $-4$ to $-2.2$ 
according to $\nu$, it is nearly constant when 
$0.001 \le \nu \le 0.003$, which is obtained by the 
proper resolution and closer to that of Saffman than that of 
Batchelor. (See the energy spectrum in Figure 4d) 
The power $p$ is obtained as 
$-0.171$ by the two cases $\nu = 0.001, 0.003$ and the energy 
averaged in the time interval $7 \le t \le 10$, 
which is close to but 
slightly larger than Saffman's value $-0.133$. 
Comparing \Eq{n1} with \Eq{t27}, relations between $p, q$ and $\ga$ are 
obtained as 
\begin{equation}
p= \fr{\ga (1-\ga)}{\ga-2},
\label{n2} \end{equation}
\begin{equation}
q= \fr{4-\ga}{\ga-2}.  
\label{n3} \end{equation}
Although $\ga$ varies from 0.333 to 1.333 from the numerical data of 
$q$, the plausible value of $p=-0.171$ leads to $\ga=0.614$ or 
0.557. From \Eq{t22} we obtain $\et \approx \nu^{1-\ga}$ and the power 
lies between 0.386 to 0.443. The obtained power locates between 
the KLB value 0 and the Saffman's value 2/3. 

The temporal evolution of the inviscid 
invariant $\Om_n$ and its dissipation rate $\et_n$ 
is examined and shown in Figures 5a,b for 
$\nu=0.0003$ and even $n$ between 4 and 20. For odd $n$, $\Om_n$ 
oscillates irregularly around zero and no clear dependence 
on $\nu$ or $n$ is observed. The adopted numerical method 
does not conserve $\Om_n$ even if $\nu=0$ because of truncation 
in the Fourier space, although energy and enstrophy are conserved. 
In fact, oscillations appear in $\Om_n$ in the case of the smallest 
viscosity and large $n$. However, we analyze the data of $\et_n$ 
since the deviation is not monotonic and the decay dominates 
the oscillation for the most cases. 

Similar to the $\et$, the following asymptotic form of $\et_{n}$ 
for even $n$ is assumed:
\begin{equation}
\eta_{n} \propto (\nu / E )^{p_{n}}(t+ t_0)^{q_{n}}. 
\label{n4} \end{equation}
Figure 3a shows the powers $q_{n}$ versus $n$ for the four cases. 
If there is an $n$ dependence of $q_{n}$ predicted by conformal field 
theory like 
\begin{equation}
q_{n} = q_0 + n q_1, 
\label{n5} \end{equation}
we may estimate them by the data shown in Figure 7. 
Table 3 shows estimated values of $q_0$ and $q_1$. 
Constants $q_0$ are within the values of \Eq{ca8} or \Eq{ca9} 
predicted by category A. 
However, for $q_1$, numerical values are smaller than theoretical ones. 

The other parameter $p_n$, denoted by 
\begin{equation}
p_{n} = p_0 + n p_1, 
\label{n6} \end{equation}
is also estimated by plotting 
$\log C_n /\log (\nu/E)$ where $C_n$ is 
$\eta_{n} / (t+t_0)^{q_{n}}$. 
We find negative values, 
between -0.422 and -0.265, of $p_0$ and, 
between -0.0145 and -0.00525, of $p_1$ the latter is very small. 
Since the constants $C_n$ include a constant proportional to 
$n^2$, the degree of coincidence is unsatisfactory. 

For skewness and flatness, we should determine what kind of 
variables is to be measured. We choose 
\begin{equation}
S= -\fr{2<u_x \om_x^2>}{<u_x^2>^{1/2}<\om^2_x>}
\label{n7} \end{equation}
and
\begin{equation}
F= \fr{<\om_x^4>}{<\om^2_x>^2}
\label{n8} \end{equation}
since they show rapid growth in the initial stage 
$t \approx 2$. 
The former is related with production of palinstrophy. 
In Figure 7a,b maxima of $S$ and $F$ versus $\nu$ at the initial 
increase are plotted. We observe that both $S$ and $F$ increases 
as $\nu$ decreases, as is observed in many experiments of 
three-dimensional turbulence \cite{Va}. 
Intermittency of turbulence is responsible for 
the tendency that skewness and flatness increase as Reynolds number 
becomes large. 
We observe that $S \approx Re^{0.117}$ and $F \approx Re^{0.222}$. 
On the other hand, the predicted feature for $S$ and $F$ is the 
opposite, i.e., the decreases as Reynolds number increases. 
Actually this prediction contradicts with observations so far. 
Since the behaviour of higher order statistics is strongly dependent 
on what is the measured physical quantity, there may exist 
observables faithful to the present prediction. 

\section{ Conclusions and Discussions }

It is shown that the extended theory of two-dimensional turbulence 
characterized by the energy exponent and the viscous scale index 
can clarify relations between classical and conformal field theories. 
New classes of CFT solutions for two-dimensional turbulence 
are obtained and its statistical predictions are explored. 
Many of the obtained exponents of the energy spectrum lie between 
or close to the reliable values $-3$ and $-4$. 
The dependence of Casimir invariants on the energy, enstrophy and 
viscosity is shown and used to predict its decay rate. 
Dependence of skewness and flatness on Reynolds number 
are also revealed and the predicted tendency is contrary to 
intermittency effects in three-dimensional turbulence. 
This may imply that the quantities to be measured are not 
the usual ones like $\pd\om/\pd x$. 

There are some possibilities for further study of higher order 
quantities. For flatness, behaviors of other quantities, e.g., 
$\psi$, $\om$, can be investigated if the integration time is 
so large that isolated coherent vortices emerge. 
To perform more precise 
numerical investigation on $<\om^n>$, we can invoke the Zeitlin's 
finite-mode system, which was studied numerically in detail 
by Hattori \cite{Ha} without viscosity. 
In this system the efficient FFT method is not available and 
its simulation requires much CPU time for integration. 

\section{ Acknowledgements }

Numerical simulation was performed by IBM RS/6000 
workstations in International Center for Elementary Particle Physics, 
University of Tokyo, 
which are supplied for a partnership program with IBM Japan, ltd. 
This work is also supported by Grant-in-Aid for Scientific Research 
from the Ministry of Education, Science and Culture in Japan. 

\section{ Figure Captions }

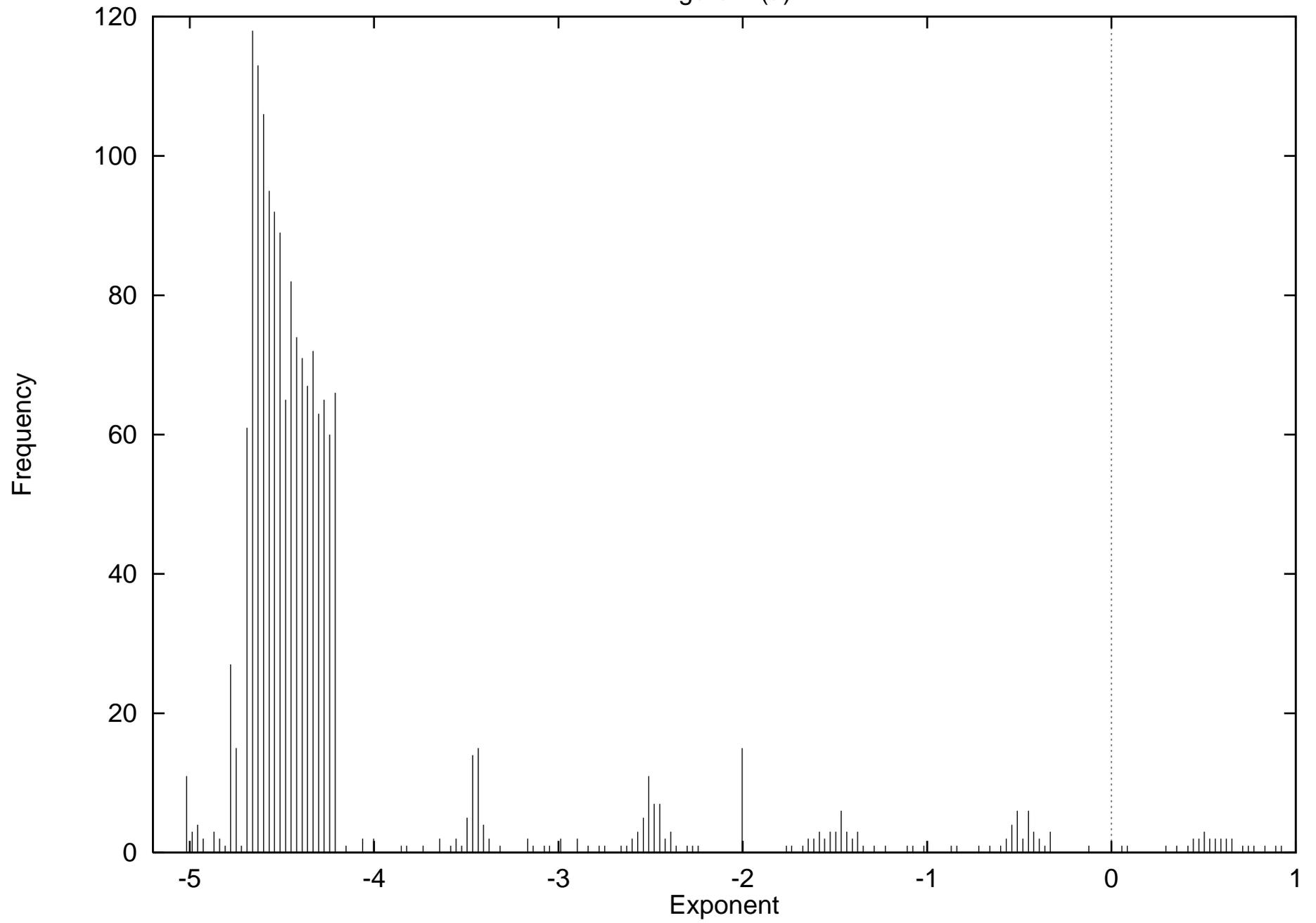
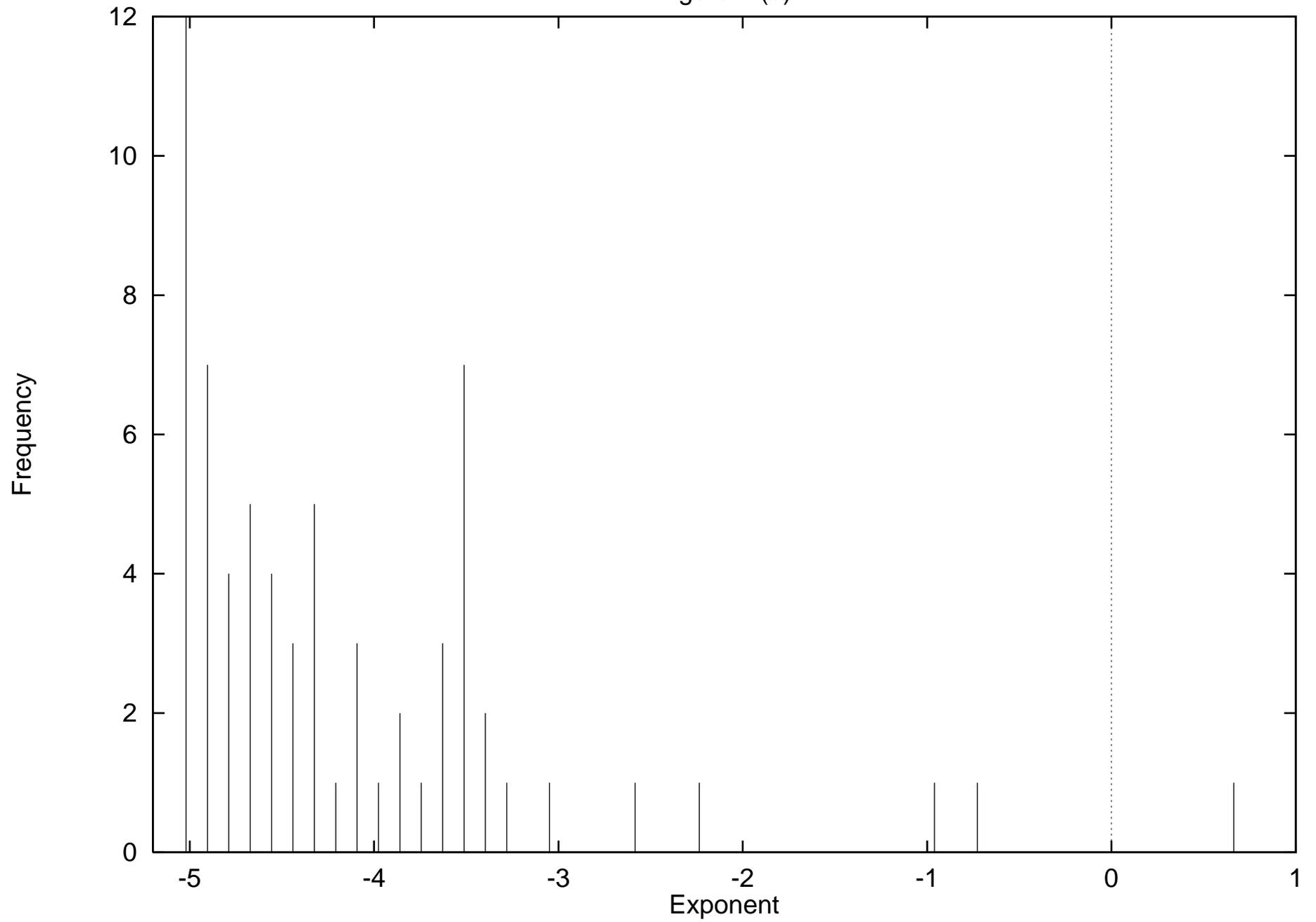
Figure 1. Histogram of the CFT solutions
versus energy exponent, by using
KLB condition (a), and Saffman condition (b).
In both of the case, we assume that the one point function
is not vanishing.

\noindent
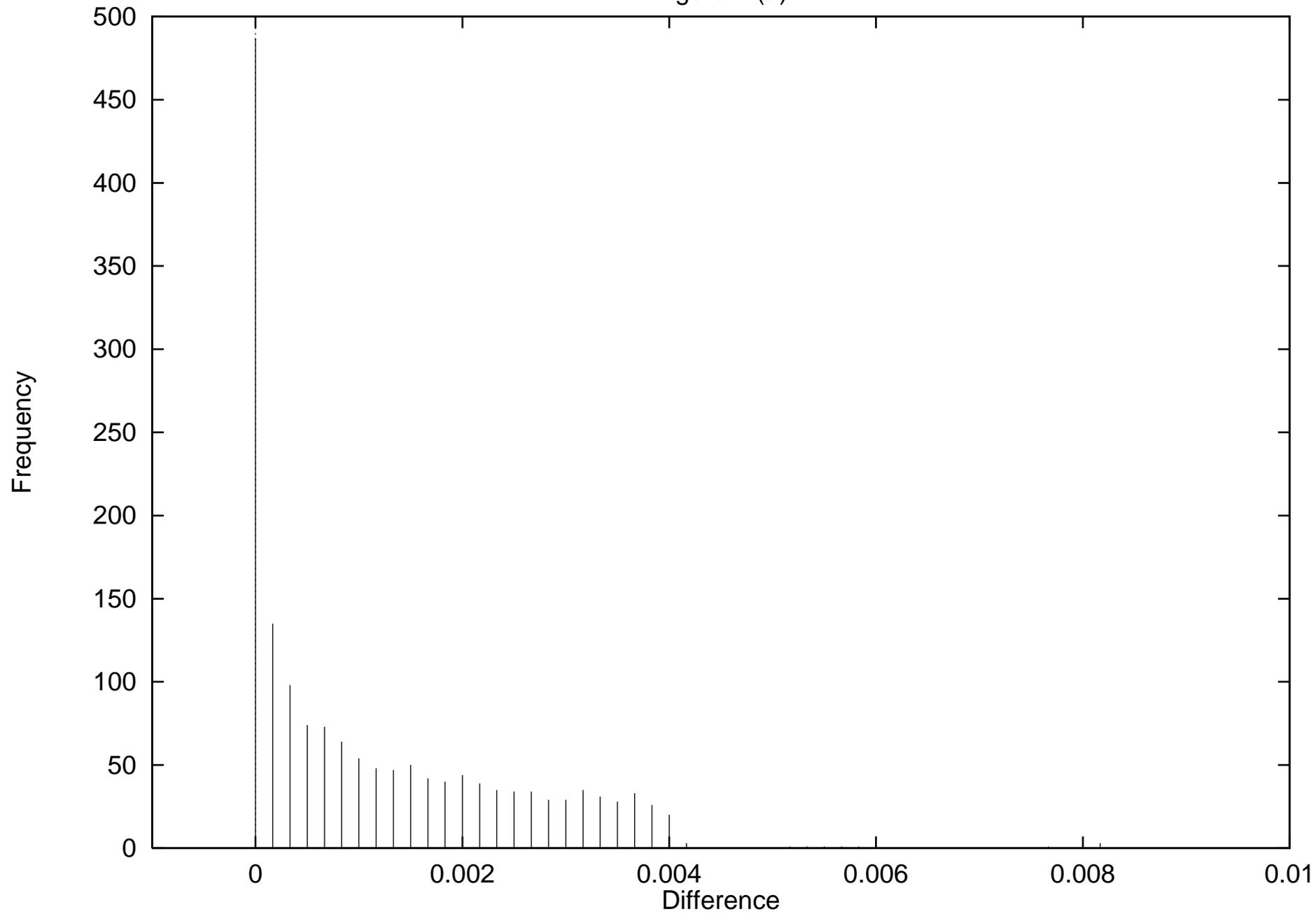
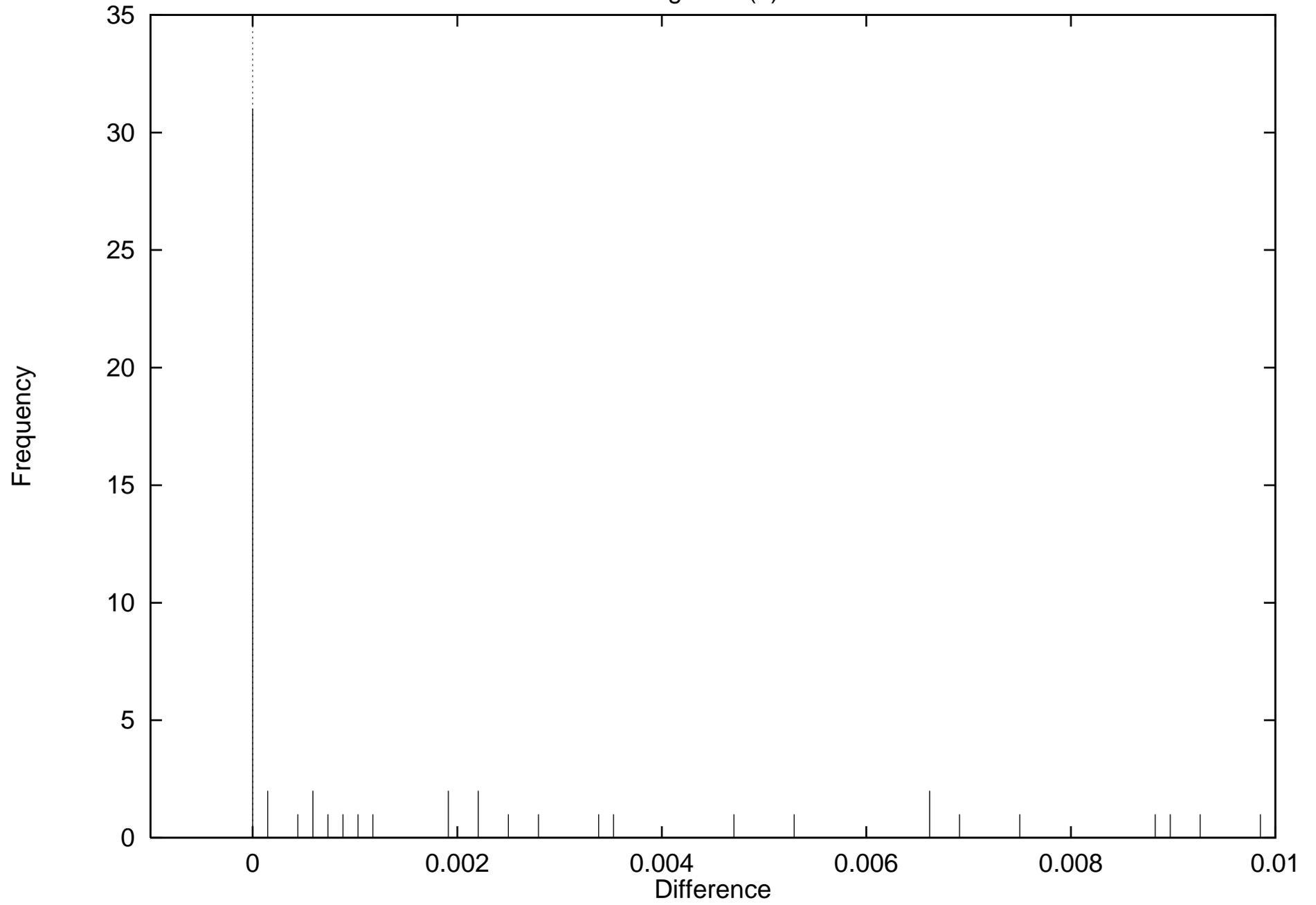
Figure 2. Histograms of the difference between
the dimension of $\phi$ and the lowest dimension of
the system. We use KLB condition (a) and
Saffman condition (b).  Here we removed datum of category B.

\noindent
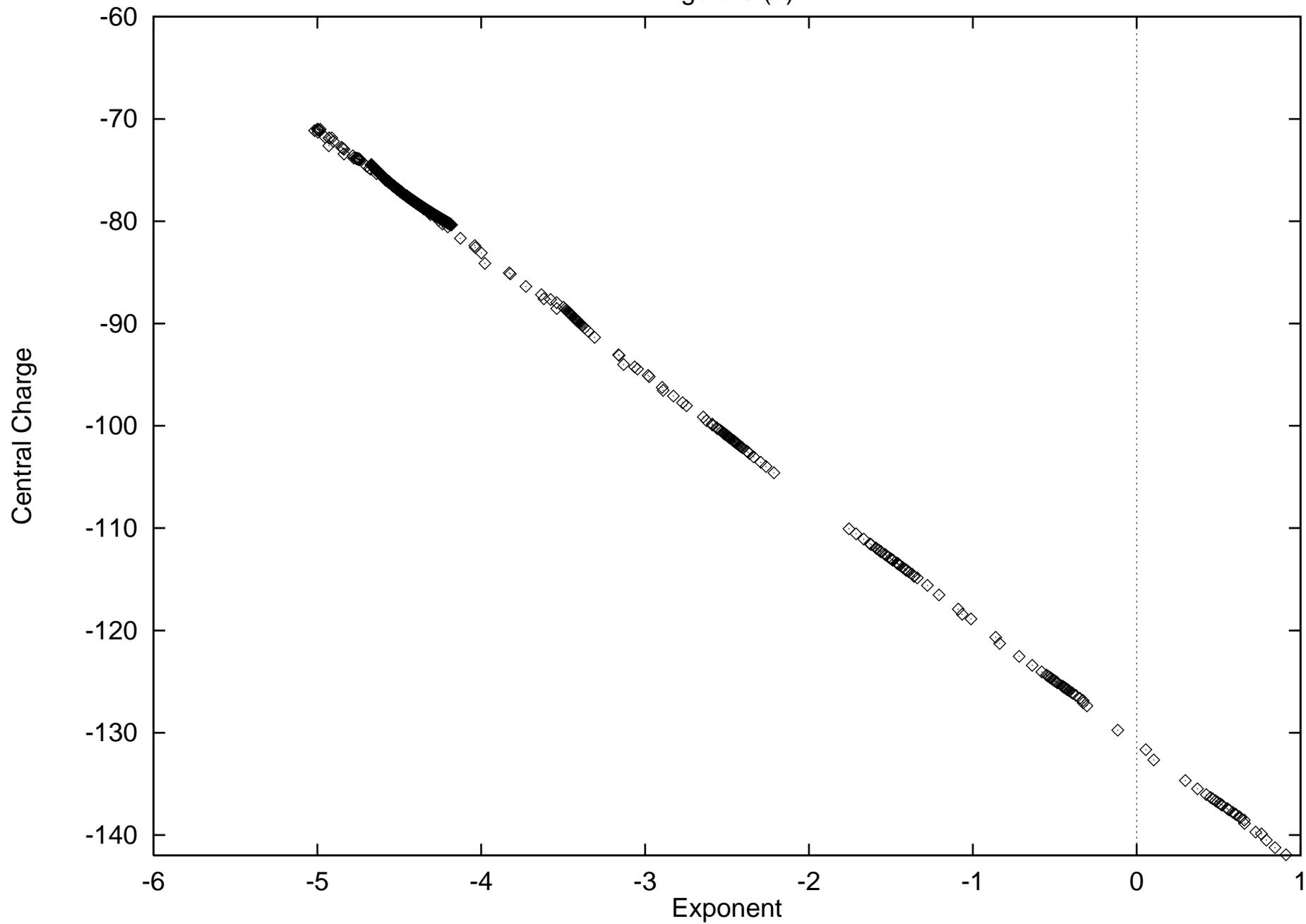
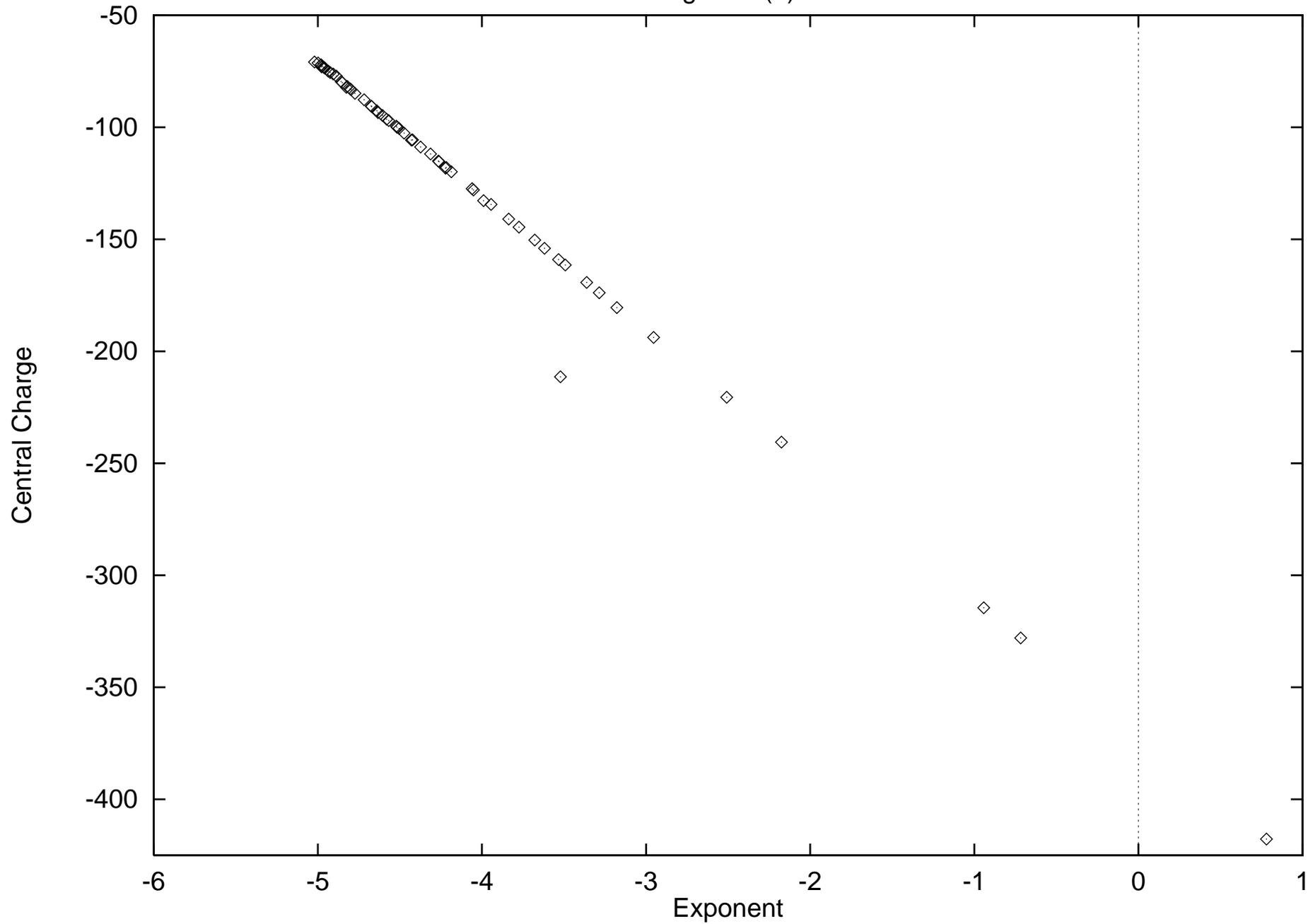
Figure 3.  Relation between the central charge and
and energy exponent. We use KLB condition (a) and
Saffman condition (b).

\noindent
Figure 4. Temporal evolution of the energy (a), enstrophy (b) and 
enstrophy dissipation rate (c) and energy spectrum (d) at $t=10$. 
Four cases with the 
viscosity $\nu=0.0001, 0.0003, 0.001, 0.003$ are drawn by 
solid, dashed, dotted and dot-dashed lines. 
A dashed straight line in (d) shows the $k^-4$ energy spectrum. 

\noindent
Figure 5. Temporal evolution of $\Om_n$ (a) and $\et_n$ (b) 
for the case $\nu=0.0003$. 
Values for $n=4, 6, 8, \cdots 20$ are denoted in the order from below. 

\noindent
Figure 6. Numerically obtained values of the powers $q_{n}$  
in \Eq{n5}. 
Cases $\nu =0.0001$, $0.0003$, $0.001$, $0.003$ are denoted by 
a circle, square, diamond and cross respectively. 

\noindent
Figure 7. Numerically obtained maxima in the initial growth of 
skewness (a) and flatness (b) defined in \Eq{n7} and in \Eq{n8} 
versus $Re$.

\section{Table Captions }

Table 1. First few solutions for CFT turbulence. 
(a) KLB condition (with non-vanishing one-point function)
\begin{center}
\begin{tabular}{|c|c|c|c|c|}
\hline
(p,q) & $\psi$ & $\phi$ & $\chi$ & $\alpha$\\
\hline
(33,2) & $\Delta_{10,1}\sim-3.00$ &$\Delta_{17,1}\sim-3.64$ &$\Delta_{16,1}\sim-3.64$ &-3.73\\ 
 \hline 
(43,3) & $\Delta_{12,1}\sim-3.01$ &$\Delta_{15,1}\sim-3.09$ &$\Delta_{14,1}\sim-3.10$ &-4.84\\ 
 \hline 
(44,3) & $\Delta_{18,1}\sim-2.99$ &$\Delta_{15,1}\sim-3.18$ &$\Delta_{14,1}\sim-3.18$ &-4.61\\ 
 \hline 
(46,3) & $\Delta_{20,1}\sim-2.99$ &$\Delta_{15,1}\sim-3.35$ &$\Delta_{16,1}\sim-3.34$ &-4.28\\ 
 \hline 
(52,3) & $\Delta_{25,1}\sim-3.00$ &$\Delta_{17,1}\sim-3.85$ &$\Delta_{17,1}\sim-3.85$ &-3.31\\ 
 \hline 
(67,3) & $\Delta_{36,1}\sim-3.00$ &$\Delta_{23,1}\sim-5.09$ &$\Delta_{22,1}\sim-5.09$ &-0.84\\ 
 \hline 
(77,3) & $\Delta_{43,1}\sim-3.00$ &$\Delta_{25,1}\sim-5.92$ &$\Delta_{25,1}\sim-5.92$ &0.84\\ 
 \hline 
(100,3) & $\Delta_{11,1}\sim-4.10$ &$\Delta_{21,1}\sim-6.70$ &$\Delta_{31,1}\sim-7.80$ &-2.00\\ 
 \hline 
(121,3) & $\Delta_{12,1}\sim-4.61$ &$\Delta_{23,1}\sim-7.73$ &$\Delta_{34,1}\sim-9.34$ &-2.00\\ 
 \hline 
\end{tabular}
\end{center}

(b) Saffman condition (with non-vanishing one-point function)
\begin{center}
\begin{tabular}{|c|c|c|c|c|}
\hline
(p,q) & $\psi$ & $\phi$ & $\chi$ & $\alpha$\\
\hline
(97,4) & $\Delta_{12,1}\sim-4.03$ &$\Delta_{23,1}\sim-5.56$ &$\Delta_{24,1}\sim-5.57$ &-3.99\\ 
 \hline 
(74,5) & $\Delta_{12,1}\sim-3.08$ &$\Delta_{15,1}\sim-3.22$ &$\Delta_{14,1}\sim-3.21$ &-4.91\\ 
 \hline 
(86,5) & $\Delta_{23,1}\sim-3.33$ &$\Delta_{17,1}\sim-3.81$ &$\Delta_{17,1}\sim-3.81$ &-4.67\\ 
 \hline 
(139,5) & $\Delta_{43,1}\sim-4.38$ &$\Delta_{27,1}\sim-6.45$ &$\Delta_{27,1}\sim-6.45$ &-3.62\\ 
 \hline 
(284,5) & $\Delta_{95,1}\sim-7.28$ &$\Delta_{57,1}\sim-13.70$ &$\Delta_{57,1}\sim-13.70$ &-0.72\\ 
 \hline 
(131,6) & $\Delta_{32,1}\sim-3.79$ &$\Delta_{21,1}\sim-4.96$ &$\Delta_{22,1}\sim-4.97$ &-4.22\\ 
 \hline 
(117,7) & $\Delta_{22,1}\sim-3.28$ &$\Delta_{17,1}\sim-3.69$ &$\Delta_{16,1}\sim-3.69$ &-4.72\\ 
 \hline 
(299,8) & $\Delta_{14,1}\sim-5.20$ &$\Delta_{27,1}\sim-8.13$ &$\Delta_{38,1}\sim-8.85$ &-3.52\\ 
 \hline 
(196,9) & $\Delta_{77,4}\sim-3.78$ &$\Delta_{109,5}\sim-4.96$ &$\Delta_{87,4}\sim-4.96$ &-4.22\\ 
 \hline 
\end{tabular}
\end{center}

(c) KLB condition (with vanishing one-point function)
\begin{center}
\begin{tabular}{|c|c|c|c|c|}
\hline
(p,q) & $\psi$ & $\phi$ & $\chi$ & $\alpha$\\
\hline
(21,2) & $\Delta_{4,1}\sim-1.14$ &$\Delta_{7,1}\sim-1.86$ &$\Delta_{10,1}\sim-2.14$ &-3.57\\ 
 \hline 
(25,3) & $\Delta_{11,1}\sim-1.40$ &$\Delta_{9,1}\sim-1.60$ &$\Delta_{9,1}\sim-1.60$ &-4.60\\ 
 \hline 
(26,3) & $\Delta_{5,1}\sim-1.31$ &$\Delta_{9,1}\sim-1.69$ &$\Delta_{9,1}\sim-1.69$ &-4.23\\ 
 \hline 
(55,6) & $\Delta_{14,1}\sim-1.18$ &$\Delta_{9,1}\sim-1.82$ &$\Delta_{10,1}\sim-1.80$ &-3.73\\ 
 \hline 
(62,7) & $\Delta_{13,1}\sim-1.26$ &$\Delta_{9,1}\sim-1.74$ &$\Delta_{9,1}\sim-1.74$ &-4.03\\ 
 \hline 
(67,8) & $\Delta_{28,3}\sim-1.38$ &$\Delta_{25,3}\sim-1.62$ &$\Delta_{42,5}\sim-1.62$ &-4.51\\ 
 \hline 
(71,9) & $\Delta_{32,4}\sim-1.50$ &$\Delta_{55,7}\sim-1.50$ &$\Delta_{32,4}\sim-1.50$ &-4.99\\ 
 \hline 
(87,11) & $\Delta_{16,2}\sim-1.51$ &$\Delta_{23,3}\sim-1.49$ &$\Delta_{16,2}\sim-1.51$ &-5.03\\ 
 \hline 
(91,11) & $\Delta_{14,2}\sim-1.40$ &$\Delta_{25,3}\sim-1.60$ &$\Delta_{16,2}\sim-1.59$ &-4.61\\ 
 \hline 
\end{tabular}
\end{center}

(d) Saffman condition (with vanishing one-point function)
\begin{center}
\begin{tabular}{|c|c|c|c|c|}
\hline
(p,q) & $\psi$ & $\phi$ & $\chi$ & $\alpha$\\
\hline
(17,2) & $\Delta_{6,1}\sim-1.47$ &$\Delta_{9,1}\sim-1.65$ &$\Delta_{8,1}\sim-1.65$ &-4.88\\ 
 \hline 
(41,2) & $\Delta_{4,1}\sim-1.32$ &$\Delta_{7,1}\sim-2.41$ &$\Delta_{10,1}\sim-3.29$ &-4.27\\ 
 \hline 
(46,3) & $\Delta_{27,1}\sim-1.13$ &$\Delta_{15,1}\sim-3.35$ &$\Delta_{15,1}\sim-3.35$ &-3.52\\ 
 \hline 
(46,5) & $\Delta_{13,1}\sim-1.43$ &$\Delta_{9,1}\sim-1.83$ &$\Delta_{9,1}\sim-1.83$ &-4.74\\ 
 \hline 
(53,6) & $\Delta_{12,1}\sim-1.45$ &$\Delta_{9,1}\sim-1.74$ &$\Delta_{8,1}\sim-1.72$ &-4.81\\ 
 \hline 
(65,6) & $\Delta_{17,1}\sim-1.35$ &$\Delta_{11,1}\sim-2.23$ &$\Delta_{11,1}\sim-2.23$ &-4.42\\ 
 \hline 
(91,8) & $\Delta_{41,3}\sim-1.33$ &$\Delta_{57,5}\sim-2.37$ &$\Delta_{57,5}\sim-2.37$ &-4.31\\ 
 \hline 
(131,8) & $\Delta_{62,3}\sim-1.08$ &$\Delta_{49,3}\sim-3.61$ &$\Delta_{82,5}\sim-3.61$ &-3.31\\ 
 \hline 
(79,10) & $\Delta_{39,5}\sim-1.50$ &$\Delta_{71,9}\sim-1.51$ &$\Delta_{39,5}\sim-1.50$ &-4.99\\ 
 \hline 
\end{tabular}
\end{center}

\noindent
Table 2. Numerically obtained values of the time offset $t_0$, 
the power $q$ in \Eq{n1} and $\ga$ in \Eq{n3}.

\begin{center}
\begin{tabular}{|c|c|c|c|}
\hline
$\nu$  & $t_0$  & $q$      & $\ga$   \\
\hline
0.0001 & 24.3   & $-4.03$  & $1.34$  \\
\hline
0.0003 & 16.8   & $-3.61$  & $1.23$  \\
\hline
0.001  & 3.58   & $-2.28$  & $0.438$ \\
\hline
0.003  & 1.21   & $-2.36$  & $0.529$ \\
\hline
\end{tabular}
\end{center}

\noindent
Table 3. Numerically obtained values of 
the constants $q_0$ and $q_1$ in \Eq{n5}.

\begin{center}
\begin{tabular}{|c|c|c|c|c|}
\hline
$\nu$  & 
$q_0 \ (4 \le n \le 20) $  & $q_1 \ (4 \le n \le 20)$ & 
$q_0 \ (12 \le n \le 20) $  & $q_1 \ (12 \le n \le 20)$   \\
\hline
0.0001 & $-7.41$ & $-0.326$ & $-8.64$  & $-0.245$  \\
\hline
0.0003 & $-5.84$ & $-0.0997$& $-3.43$  & $-0.245$  \\
\hline
0.001  & $-2.95$ & $-0.126$ & $-1.35$  & $-0.225$  \\
\hline
0.003  & $-1.86$ & $-0.451$ & $-1.05$  & $-0.503$  \\
\hline
\end{tabular}
\end{center}

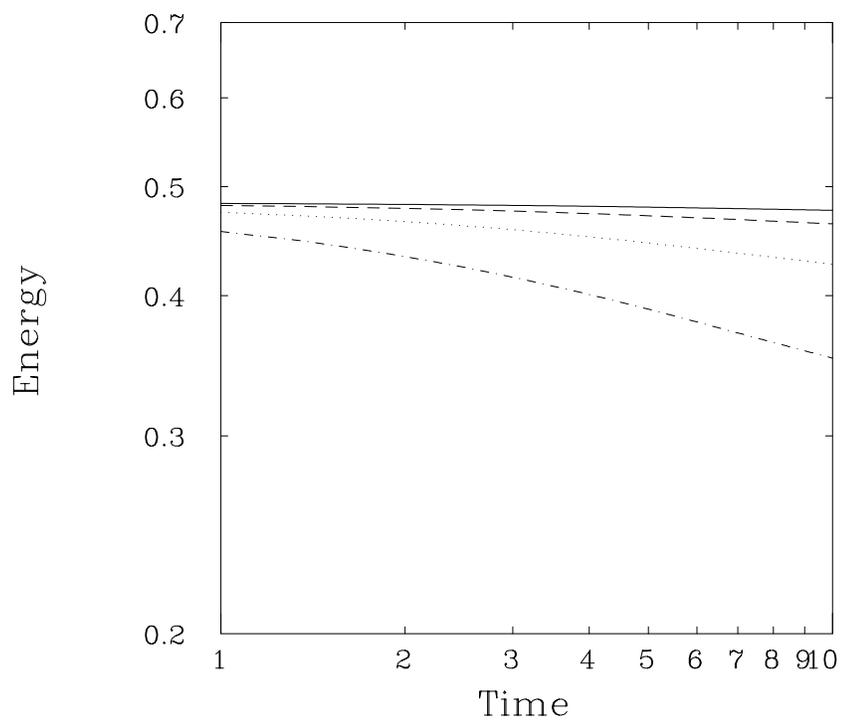
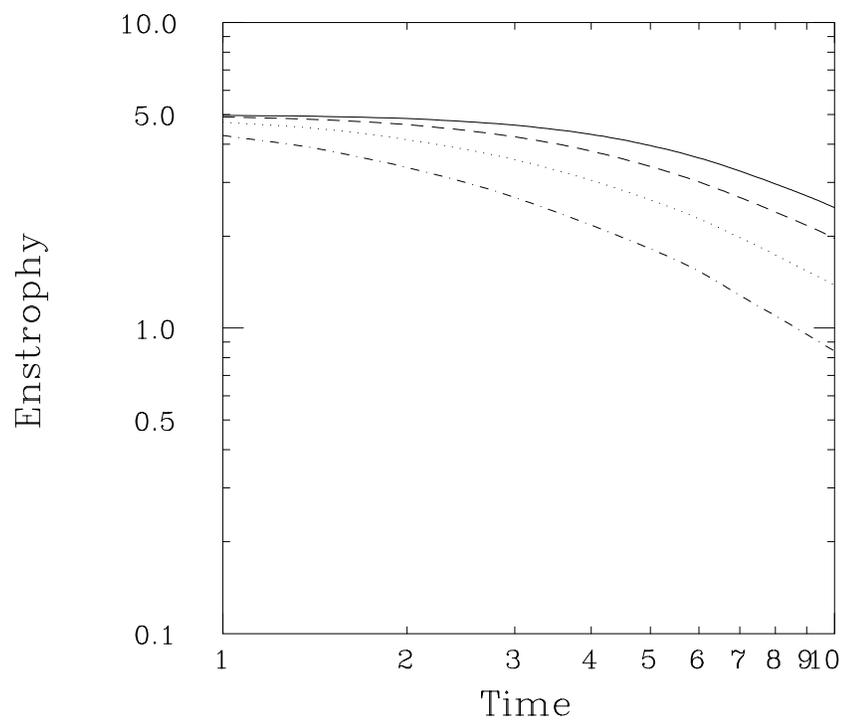
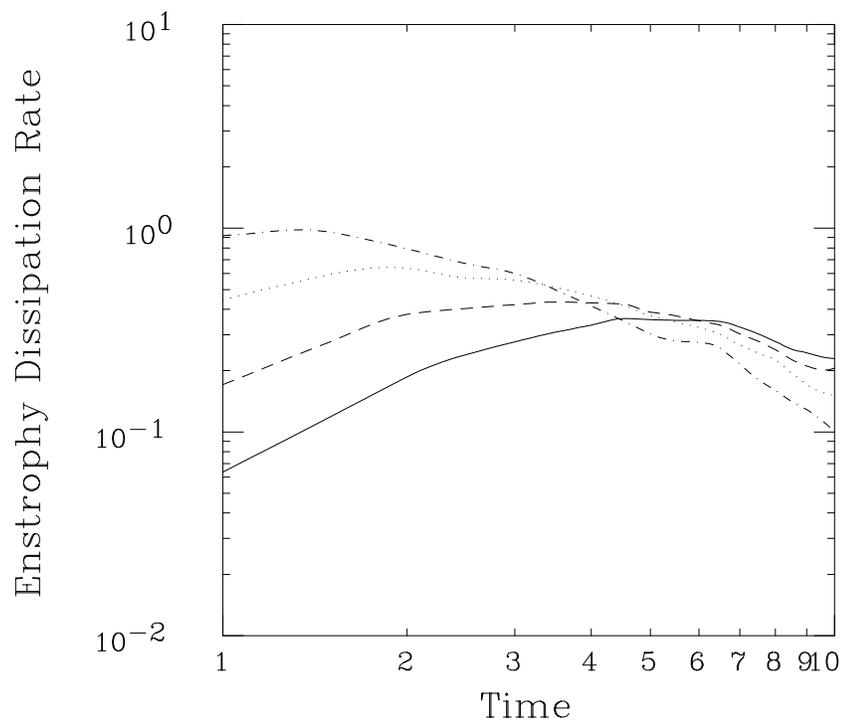
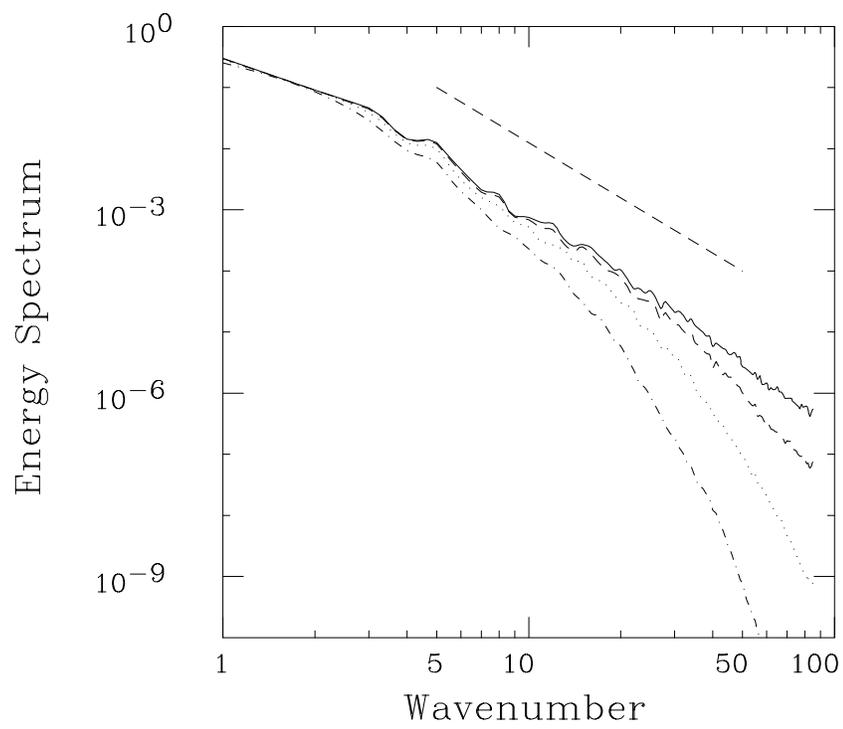
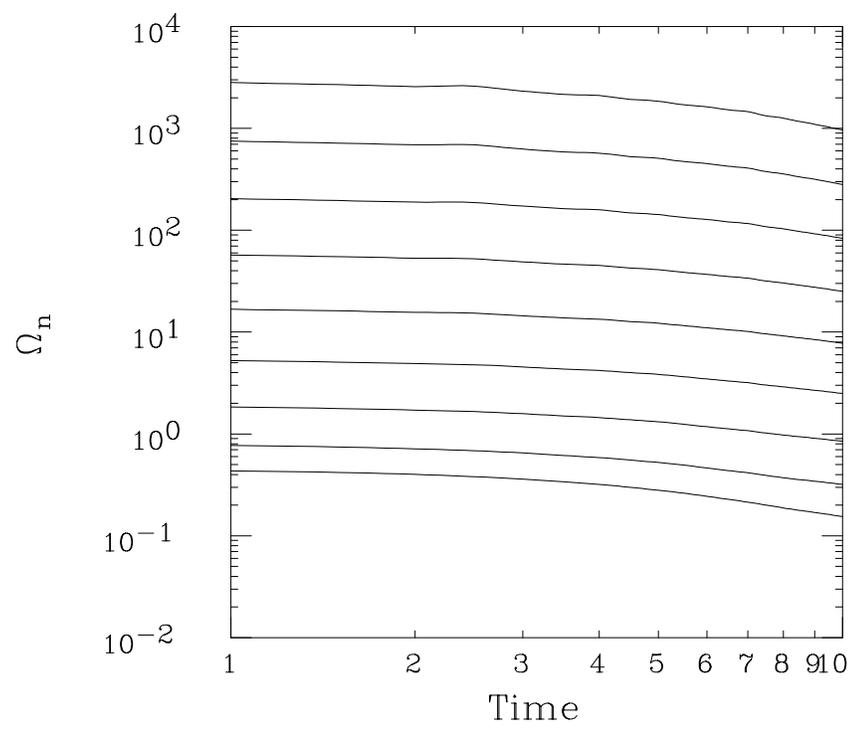
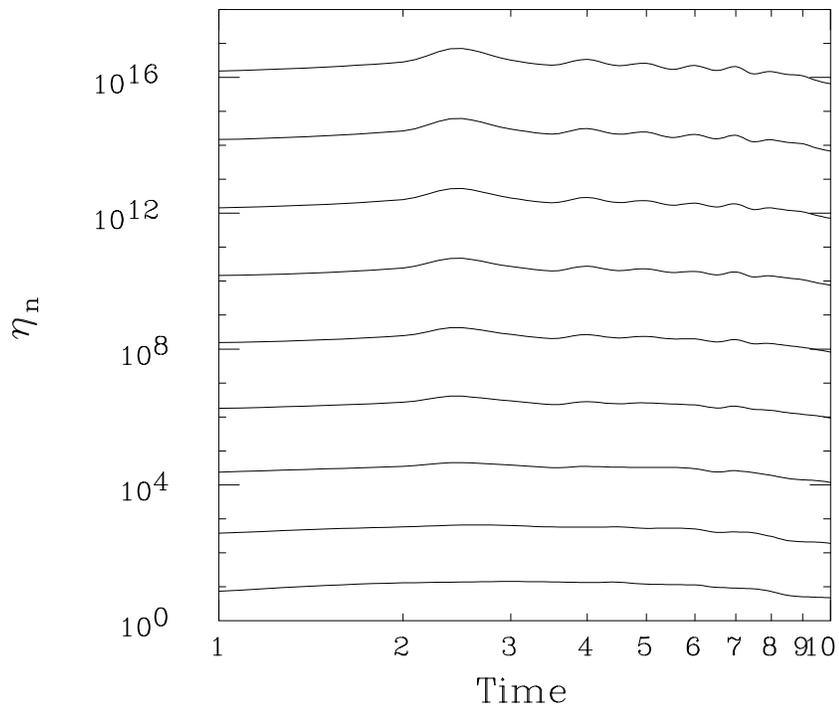
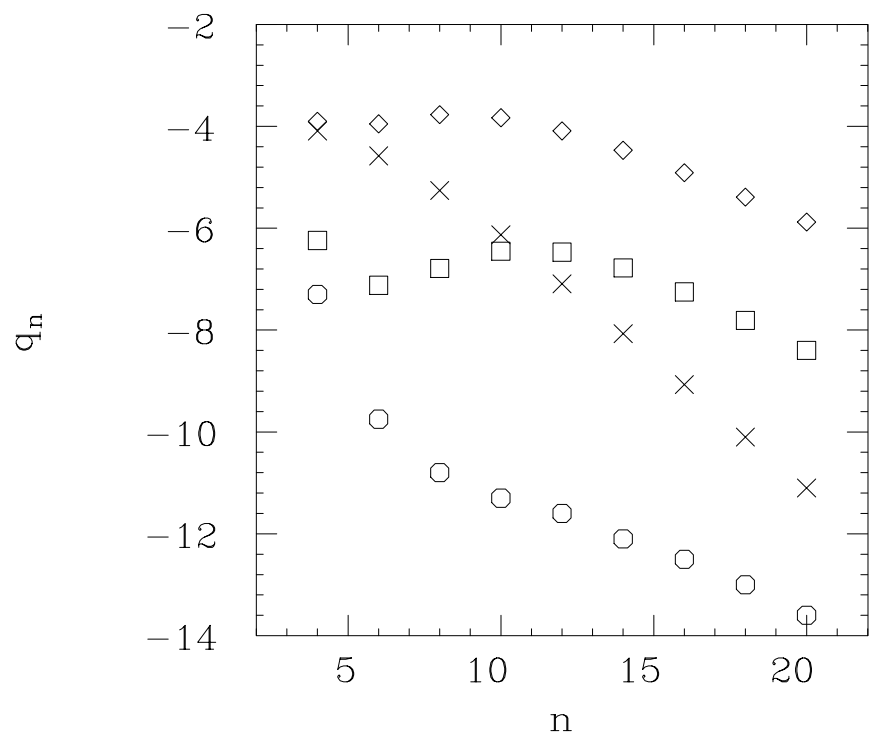
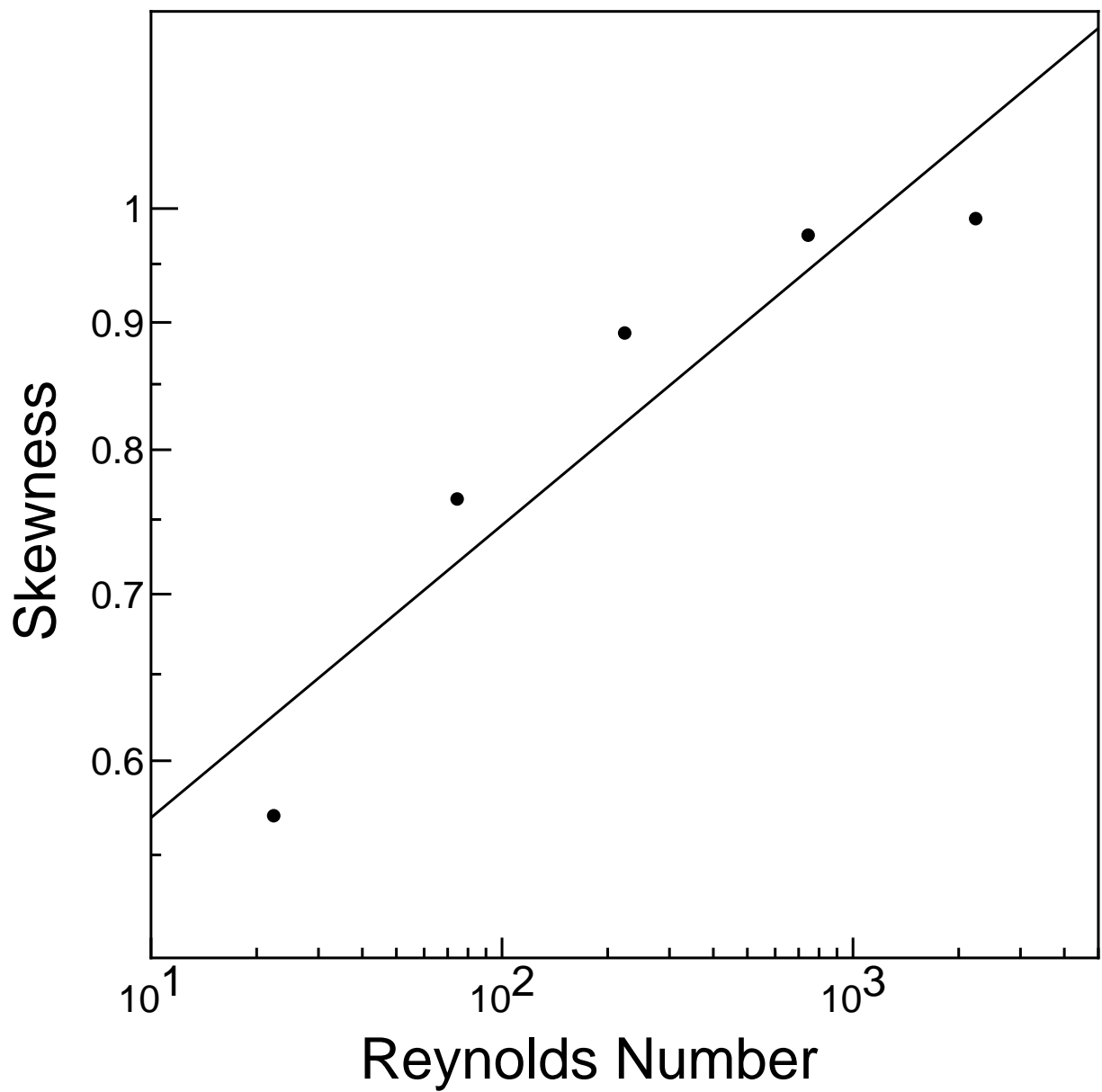
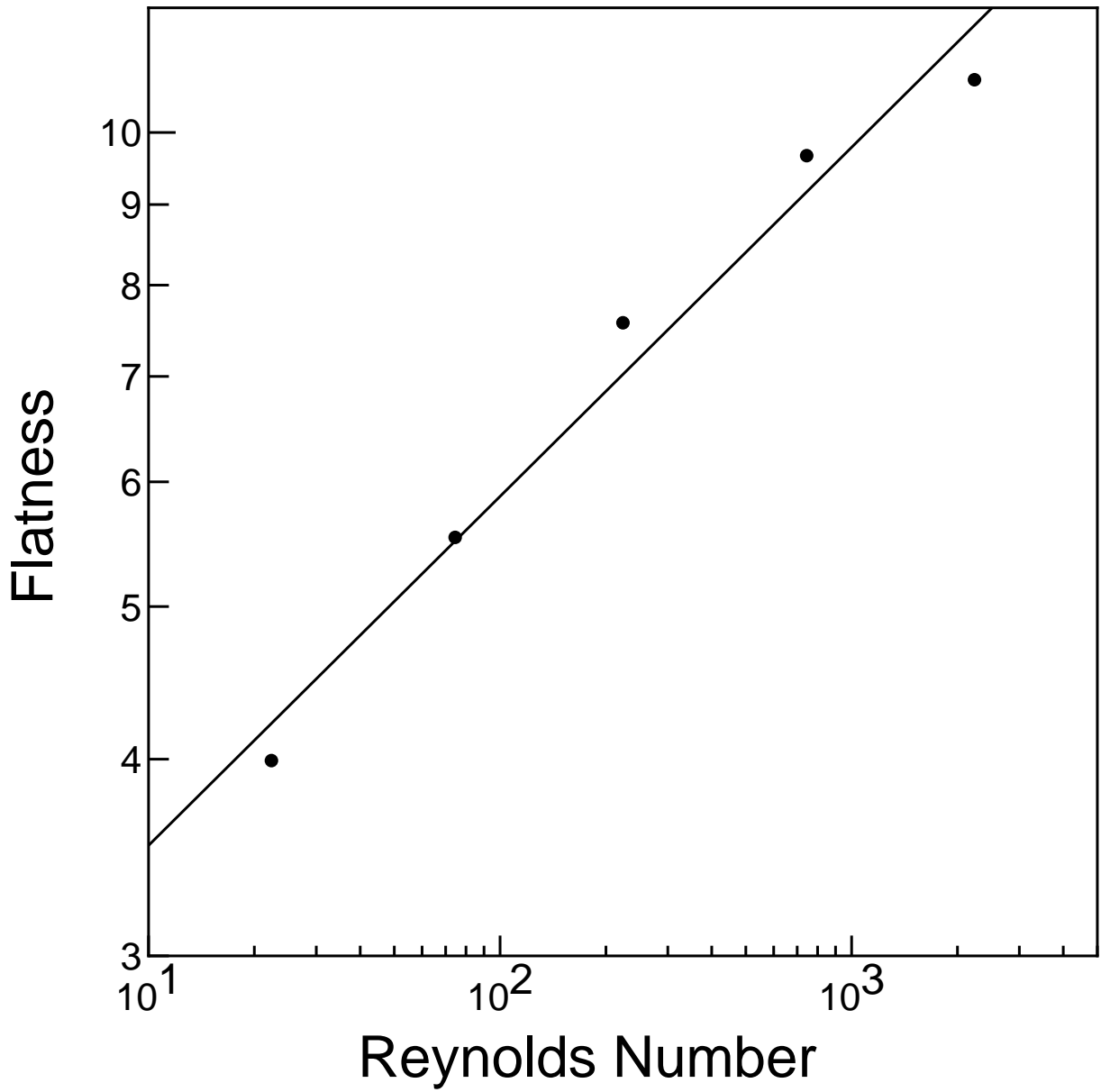
\end{document}